\documentclass[a4paper,twocolumn,aps,eqsecnum,nofootinbib,notitlepage,superscriptaddress]{revtex4}
\usepackage{graphicx}% Include figure files
\usepackage{bm}% bold math
\usepackage{amsmath}
\usepackage{amssymb}
\usepackage{amsfonts}
\usepackage{ulem}
\usepackage[dvips]{color}
\newcommand{\half}{\frac{1}{2}}

\newcommand{\der}{\partial}
\newcommand{\bra}{\langle}
\newcommand{\ket}{\rangle}

\newcommand{\Tr}{\mbox{\rm Tr}}
\newcommand{\dsl}{\partial\kern-0.55em\raise 0.14ex\hbox{/}}
\newcommand{\bfk}{\bm{k}}
\newcommand{\bfl}{\bm{l}}

\newcommand{\bfP}{\bm{P}}

\begin{document}

\preprint{KYUSHU-HET-134}

\title{Wilsonian renormalization group analysis of nonrelativistic
three-body systems without introducing dimerons}

\author{Koji Harada}
\email{harada@phys.kyushu-u.ac.jp}
\affiliation{Department of Physics, Kyushu University\\
Fukuoka 812-8581 Japan}
\author{Hirofumi Kubo}
\email{kubo@higgs.phys.kyushu-u.ac.jp}
\affiliation{Synchrotron Light Application Center, Saga University \\
1 Honjo, Saga 840-8502, Japan}
\author{Issei Yoshimoto}
\email{yoshimoto@higgs.phys.kyushu-u.ac.jp}
\affiliation{Department of Physics, Kyushu University\\
Fukuoka 812-8581 Japan}

\date{\today}

\begin{abstract}
 Low-energy effective field theory describing a nonrelativistic
 three-body system is analyzed in the Wilsonian renormalization group
 (RG) method. No effective auxiliary field (dimeron) that corresponds to
 two-body propagation is introduced. 

 The Efimov effect is expected in the case of an infinite two-body
 scattering length, and is believed to be related to the limit cycle
 behavior in the three-body renormalization group equations (RGEs).  If
 the one-loop property of the RGEs for the nonrelativistic system
 without the dimeron field, which is essential in deriving RGEs in the
 two-body sector, persists in the three-body sector, it appears to
 prevent the emergence of limit cycle behavior.  We explain how the
 multi-loop diagrams contribute in the three-body sector without
 contradicting the one-loop property of the RGEs, and derive the correct
 RGEs, which lead to the limit cycle behavior.  The Efimov parameter,
 $s_{0}$, is obtained within a few percent error in the leading
 orders. We also remark on the correct use of the dimeron formulation.

 We find rich RG-flow structure in the three-body sector. In particular,
 a novel nontrivial fixed point of the three-body couplings is found
 when the two-body interactions are absent. We also find, on the
 two-body nontrivial fixed point, the limit cycle is realized as a loop
 of finite size in the space of three-body coupling constants when terms
 with derivatives are included.
\end{abstract}

%\pacs{}
%\keywords{}

\maketitle

\newcommand{\comment}[1]{{\color{red}\tiny #1}}

%\tableofcontents

\section{Introduction}

%\comment{physics in 3-body systems}

Nonrelativistic three-body systems have many interesting features. There
have been long-standing discrepancies and anomalies between theoretical
calculations and the experiments for several physical quantities in the
three nucleon systems. (See Ref.~\cite{KalantarNayestanaki:2011wz} for a
review.) So-called Borromean nuclei, three-body bound states, any two of
whose constituents do not form a bound state, are known to exist. (See
Ref.~\cite{Zhukov:1993aw} for a review.)  V.~Efimov showed that there
are infinitely many three-body bound states with the ratios of the
binding energies of the subsequent bound states being a universal
constant, $E_{n+1}/E_{n}\equiv e^{-2\pi/s_{0}}=1/515.03\cdots$, when the
two-body scattering length is infinite~\cite{Efimov:1970zz,
Efimov:1971zz}. This ``Efimov effect'' has recently been attracting much
attention because of the advancing experimental control over the very
cold atomic systems by using Feshbach resonance.

%\comment{at low-energies, use EFT}

At low energies, the internal degrees of freedom of the particles (i.e.,
nucleons, nuclei and atoms) become irrelevant and the system is well
described by an effective field theory (EFT), in which only the
particles without internal structure interact locally. EFT is an
efficient, model-independent approach and admits a systematic
improvement of the description.

%\comment{Efimov effect in the light of RGEs}

Bedaque, Hammer, and van Kolck~\cite{Bedaque:1998km} first noticed that
the Efimov effect is related to the RG limit cycle behavior of the
three-body nonderivative coupling of the EFT. RG analysis has played an
important role in recognizing that the Efimov effect is a new kind of
universal phenomena in the three-body
systems~\cite{Braaten:2004rn}. There are several papers which deal with
the Efimov effect in the light of nonperturbative RG analysis.  For a
recent review, see Ref.~\cite{Hammer:2011kg}.

%\comment{in the literature...}

In their analysis, as well as subsequent studies by other authors,
the so-called dimeron field~\cite{Kaplan:1996nv}, an auxiliary effective
field which represents two-body propagation, is exploited. To our best
knowledge, there is no literature in which the connection between the
limit cycle behavior and the Efimov effect is established without a
dimeron field.  One might wonder if the limit cycle behavior can be
obtained without introducing a dimeron field.

%\comment{problems with dimeron field}

Although dimeron is a useful device, its introduction needs some
care. It is difficult to establish the relations between a set of the
coupling constants in the theory with dimeron and that in the original
theory because of a nontrivial factor in the path-integral measure
arising from the field transformation, an awkward contribution which
depends the regularization of how to define a functional
determinant. (In Appendix~\ref{sec:appendix:dimeron} we explain how the
additional contribution from the Jacobian affects the relations among
the couplings.) It is thus easier to think the theory with dimeron as
another EFT, the couplings of which are to be determined by a matching
procedure, than to make a direct connection between the two theories.

%\comment{In this paper, ... life without dimeron}

In this paper, we perform an RG analysis for a nonrelativistic
three-body system of a single (complex) scalar field without introducing
the dimeron. (It is straightforward to extend our analysis to the
fermionic case.) The RGEs are derived and the limit cycle behavior of
the three-body coupling is identified for the first time in this
formulation. We also find that the usual formulation with dimeron field
misses a certain contribution, which seems essential for obtaining the
Efimov parameter $s_{0}$ within a few percent error.

%\comment{nonrelativistic feature}

It is well-known that nonperturbative RGE has the ``one-loop'' property,
i.e., the running of the coupling constants is determined by a sum of
one-loop diagrams with the propagator being the full one. In a
nonrelativistic system, because of the particle number conservation and
the absence of the antiparticle, there is no dressing for the
propagator. This nonrelativistic feature divides the space of states
into sectors labeled by particle numbers, and makes the RGEs in the
two-body sector very simple: one only needs to consider actual one-loop
diagrams that satisfy the particle number conservation, with the free
propagator.

%\comment{the ``one-loop'' property puzzle}

If this one-loop property (with free propagator) of the RGEs persists in
the three-body sector, there appears to be no chance for the RGEs to
have a limit cycle solution. This is a puzzling situation: on one hand,
the Efimov effect implies the limit cycle behavior when the two-body
couplings are tuned to the critical values, on the other hand, the
general consideration given above implies the one-loop property, which
seems to prevent the limit cycle behavior. It is one of our main results
to explain how the one-loop property of the RGEs is consistent with
contributions from multi-loop diagrams in the three-body sector. The
resulting RGEs do allow limit cycle behavior.

%\comment{rich structure}

We find very rich structures in the RG flow of the three-body coupling
constants. In the leading-order calculation, in which only the
nonderivative three-body coupling is included, we find a nontrivial
fixed point on the two-body trivial fixed point. It implies the
existence of a three-body strong coupling phase with the phase boundary
on which the nontrivial fixed point resides. Note that such a nontrivial
fixed point would only be revealed in the formulation without dimerons.

%\comment{limit cycle realized as a loop}

Extending the space of three-body operators up to including the terms
with two derivatives, we find that, on the two-body nontrivial fixed
point, the limit cycle is realized as a loop of finite size, and a
nontrivial fixed point associated to it.

%\comment{effects of nonzero values of effective range and off-shell parameter}

We also emphasize that the nonzero values of the two-body effective
range and the off-shell parameter, which does not affect the on-shell
two-body amplitude, modify the ratios of the binding energies of the
three-body bound states. Although the effects of the nonzero value of
the two-body effective range have been investigated~\cite{Efimov:1991zz,
Efimov:1993zz}, those of the off-shell parameter have never been
considered.

%\comment{comments on the literature}

Several comments on the literature are in order. Although Bedaque,
Hammer, and van Kolck~\cite{Bedaque:1998km} first showed the connection
between the RG limit cycle and the Efimov effect, it does not come as a
direct consequence of the RGE analysis, but rather from the
scale-invariant behavior of the three-body amplitude. Actually they
obtained the cutoff dependence of the coupling of the three-body
nonderivative contact interaction \textit{after} determining the Efimov
parameter.  Moroz, Floerchinger, Schmidt, and
Wetterich~\cite{Moroz:2008fy} derived the Efimov parameter in the
functional RG formulation with the dimeron field. Their leading-order
value of the Efimov parameter contains about 40\% errors. Their
numerical calculation for the full RGE shows, however, that the
resulting value of the Efimov parameter is in good agreement with the
Efimov's.

%\comment{there are other works}

There are several other papers devoted to the RG analysis of the
three-body systems. For example, Diehl, Krahl and
Schrerer~\cite{Diehl:2007xz} study the ratio of dimeron-particle
scattering length of to that of particle-particle. Krippa, Walet, and
Birse~\cite{Krippa:2009vu} do a similar thing for the ratio of
dimeron-dimeron scattering length to that of particle-particle. All of
these employ dimerons in their analysis.

%\comment{structure of the paper}

The paper is organized as follows. In Sec.~\ref{sec:EFT}, we
recapitulate the importance of Wilsonian RG analysis in revealing the
physical features of few-body systems described by EFT. The one-loop
property puzzle is explained and the solution is given in
Sec.~\ref{sec:puzzle}.  Several examples are shown for the illustration.
We derive the RGEs for the coupling constant for the three-body
interaction in the leading order in derivative expansion in
Sec.~\ref{sec:derivation}. Comments on the difference between the
dimeron formulation and ours are also given. In Sec.~\ref{sec:RGflow},
we first concentrate on the case in which the two-body effective range
and the off-shell parameter are zero. The RG flows are given and the
Efimov parameter is obtained on the two-body nontrivial fixed point. The
other cases are also examined. A novel nontrivial fixed point in the
three-body coupling is identified. We examined the next-to-leading order
corrections in Sec.~\ref{sec:corrections}. The RG flow in the three-body
coupling space is presented. The limit cycle is found to be realized as
a loop of finite size in this space. We summarize the results in
Sec.~\ref{sec:summary}.

In Appendix~\ref{sec:appendix:dimeron} we illustrate a systematic way of
introducing the dimeron field in the path integral formulation. We argue
that there is a nontrivial Jacobian in the measure due to the nonlinear
change of variables. It leads to change of coupling constants which has
been neglected in the literature. In Appendix~\ref{sec:appendix:NOOL},
we demonstrate the derivation of the RGEs by using an example. In
Appendix~\ref{sec:appendix:twoloop} we identify the shell-mode
contribution of a two-loop diagram which is not obvious.  In
Appendix~\ref{sec:appendix:pole}, we explain that the pole appearing in
the RGE for the three-body coupling is due to the existence of the
two-body bound state.

\section{Effective field theory in the three-body sector and Wilsonian
 RG}
\label{sec:EFT}

In this section, we recapitulate the basic idea of
EFT~\cite{Weinberg:1978kz} and the usefulness of Wilsonian
RG~\cite{Wilson:1973jj} in the context of EFT.

\subsection{EFT}

%\comment{EFT}

Field theoretical description of nature has a hierarchical structure: at
every scale there are relevant degrees of freedom and the interactions
among them are described in terms of local operators subject to the
symmetry of the system. Even though the system is composed of composite
particles, if the momentum scale in question is smaller than the scale
of the internal structure, one may neglect the structure and describe
the system in terms of the fields corresponding to the particles. The
effects of heavier particles that are not included in the EFT are
encoded in the values of the coupling constants of local operators. Such
a field theory is called an effective filed theory (EFT). A physical
system may be described by various EFTs at various momentum scales.

%\comment{EFT is general}

The EFT description is very general: an EFT is defined by the relevant
degrees of freedom, dimensions of spacetime, and the symmetries. It
contains all possible operators with these attributes, which are
infinitely many. A single EFT may describe several different systems
that share the same attributes. Features of a specific system are
reflected in the values of the coupling constants. Thus the EFT
description is model independent.

%\comment{EFT is systematic}

The accuracy of the EFT description is controlled by power counting
rules. Power counting rules tell us the degrees of importance
(``orders'') of operators. The higher the order is, the less important
the contributions are. To a given order, one only needs to consider a
finite number of operators and a restricted set of diagrams. If the
power counting rules are consistent, counterterms are also supplied to
the order. Thus the EFT description is systematically improvable.

\subsection{Wilsonian RG reveals nonperturbative aspects of EFT}

%\comment{nonperturbative physics sometimes shows up in EFT}

In several cases, nonperturbative physics shows up in the EFT
context. It is of central importance to establish the power counting
rules for such a case to extract physical information out of EFT. An
example is the nuclear effective field theory
(NEFT)~\cite{Weinberg:1990rz, Weinberg:1991um, Epelbaum:2008ga} for the
two-nucleon system in the S waves. The existence of a bound state,
deuteron, in the spin-triplet channel, is a clear sign of
nonperturbative dynamics.

%\comment{RG transformation controls quantum fluctuations}

It is the Wilsonian, or nonperturbative, RG
analysis~\cite{Wilson:1973jj, Wegner:1972ih, Polchinski:1983gv,
Wetterich:1992yh, Morris:1993qb} that reveals properties of the
nonperturbative dynamics. The effects of quantum fluctuations are
examined scale by scale by changing the cutoff in the Wilsonian RG
analysis, without relying on the perturbative approximation.

%\comment{RGEs for EFT: NEFT as an example}

The application of such analyses to the NEFT in two-nucleon sector is
given in Refs.~\cite{Birse:1998dk, Harada:2006cw, Harada:2010ba}. The
nonperturbative feature of the system is translated into the RG language
as the existence of a nontrivial fixed point. Because of the large
anomalous dimensions, the power counting rules are modified from the
perturbative ones based on the naive dimensional analysis.

\subsection{Efimov effect in the light of RG}

%\comment{Efimov}

More than forty years ago, V.~Efimov considered the case where the
two-body scattering length is much larger than the range of the two-body
force in a nonrelativistic system, probably motivated by the large
two-nucleon scattering lengths in the S waves. He started with the
Schr\"odinger equation with a short-range two-body potential and derived
an effective three-body Schr\"odinger equation. He then noticed that
there are an infinite number of three-body bound states with the ratios
of binding energies of subsequent states being a universal constant. The
result does not depend on the details of the short-range two-body
potential.

%\comment{Efimov effect in EFT}

One may think that this phenomenon can be explained in the language of
RG in an EFT. Such a formulation would make the universal feature of the
phenomenon more transparent. Let us consider the EFT Lagrangian of a
single nonrelativistic boson of mass $M$:
\begin{eqnarray}
 {\cal L}&=&\phi^{\dagger}
  \left(
   i\der_{t}+\frac{\nabla^{2}}{2M}
  \right)\phi
  -\frac{c_{0}}{4}
  \left(
   \phi^{\dagger}\right)^{2}\phi^{2}
  \nonumber \\
 &&{}
  +\frac{c_{2}}{4}
  \left[
   \left(
    \phi^2
   \right)^{\dagger}
   \left(
    \phi
    \overleftrightarrow{\nabla}^{2}
    \phi
   \right)
   +\mbox{h.c.}
  \right]
  \nonumber \\
 &&{}
  +
  \frac{b_2}{2}
  \left[
   \left(
    \phi^2
   \right)^{\dagger}
   \phi\left(
             i\der_{t}+\frac{\nabla^{2}}{2M}
        \right)\phi
   +\mbox{h.c.}
  \right] + \cdots
  \nonumber \\
 &&{}
  -\frac{d_{0}}{36}
  \left(\phi^{\dagger}\right)^{3}\phi^{3} + \cdots,
  \label{origEFT}
\end{eqnarray}
where $\overleftrightarrow{\nabla}^2 =\overleftarrow{\nabla}^2 - 2
\overleftarrow{\nabla}\cdot\overrightarrow{\nabla} +
\overrightarrow{\nabla}^2$ and the ellipses stand for the terms with
more derivatives. Throughout this paper, we concentrate on the S waves,
so that the operators contributing to higher partial waves are not
included. We do not consider the operators which act on more than
three-body states, e.g., $(\phi^\dagger)^4\phi^4$,
$(\phi^\dagger)^5\phi^5$, etc., because we are interested in the
three-body sector and they do not contribute to the sector. Note that we
have included the so-called redundant operators such as the one on the
third line in Eq.~\eqref{origEFT}, which are necessary to renormalize
the theory off shell~\cite{Harada:2005tw}.

%\comment{BHvK}

In an important paper, Bedaque, Hammer, and van
Kolck~\cite{Bedaque:1998km} use another version of EFT with the
so-called dimeron field, $D$. In the leading order the Lagrangian is
given by
\begin{eqnarray}
 {\cal L}_{D}&=&
  \phi^\dagger
  \left(
   i\der_{t}+\frac{\nabla^{2}}{2M}
  \right)\phi
  +g_{0} D^\dagger D
  \nonumber \\
 &&{}
  +g_{1}
  \left[
   D^{\dagger}\phi^2 +(\phi^2)^{\dagger} D
  \right]
 + g_{2}
 D^{\dagger} D \phi^{\dagger} \phi
 +\cdots.
 \nonumber \\
 \label{dimeronEFT}
\end{eqnarray}
They find the limit cycle behavior for $g_{2}$ from the scale invariant
property of the three-body amplitude in the limit of an infinite
scattering length,
\begin{eqnarray}
 H(\Lambda) &\equiv& \frac{\Lambda^2g_{2}(\Lambda)}{4M g_{1}^2(\Lambda)}
  \nonumber \\
 &=&-\frac{\sin\left[s_{0}\ln(\Lambda/\Lambda_*)-\arctan(s_{0}^{-1})\right]}
  {\sin\left[s_{0}\ln(\Lambda/\Lambda_*)+\arctan(s_{0}^{-1})\right]},
\end{eqnarray}
where $s_{0}$ is called the Efimov parameter, $s_{0}=1.00624\cdots$,
$\Lambda$ is the floating cutoff, and $\Lambda_{*}$ is a constant.

%\comment{not directly from the RGE}

Note, however, that although they find the intimate connection between
the Efimov effect (as a consequence of discrete scale invariance) and
the RG limit cycle behavior, they obtain the (floating) cutoff
dependence of the coupling constant not from the RGEs. Note also that
their analytic results are given only for the case of an infinite
scattering length, and the case with a finite scattering length is
considered only numerically.

%\comment{Moroz et al. did the job. using dimeron}

More recently a field-theoretical derivation of the limit cycle behavior
directly from the RGEs is given by Moroz, Floerchinger, Schmidt, and
Wetterich~\cite{Moroz:2008fy}. They also use the dimeron field. Their
formulation admits an arbitrary momentum dependence of the coefficient
functions. In the leading-order approximation (the point-like
approximation), the RGEs are solved analytically, but in the higher
orders they are solved only numerically.

%\comment{a new derivation is desired}

Even though the use of dimeron is useful, it is just an option. The RGEs
and the limit cycle behavior should be obtained even if the dimeron
field is not introduced. Such a demonstration would provide a better
insight into the use of dimeron field.

\section{The one-loop property puzzle}
\label{sec:puzzle}

%\comment{without dimeron}

In this sections, we explain the one-loop puzzle and its solution. For
simplicity, we consider a nonrelativistic bosonic system described by
the Lagrangian~\eqref{origEFT}. The dimeron field is not introduced.

\subsection{One-loop property of nonperturbative RGEs}

%\comment{what is the ``one-loop'' property}

It is well known that nonperturbative RGEs have a kind of ``one-loop''
property. The Wegner-Houghton equation~\cite{Wegner:1972ih} for the
Wilson action $S_{\rm eff}$, for
example,
\begin{eqnarray}
 \der_{t} S_{\rm eff} &=& \frac{1}{2dt}\int_{p}'\!% \frac{d^np}{(2\pi)^n}
  \bigg\{
  %\Tr 
  \ln
  \left(
   \frac{\delta^2 S_{\rm eff}}{\delta \phi_{p}\delta \phi_{-p}}
  \right)
  \nonumber \\
 &&{}-\frac{\delta S_{\rm eff}}{\delta \phi_{p}}
  \left(\frac{\delta^2 S_{\rm eff}}
   {\delta\phi_{p}\delta\phi_{-p}}\right)^{-1}\!
  \frac{\delta S_{\rm eff}}{\delta \phi_{-p}}
  \bigg\}
  \nonumber \\
 &&{}+(\mbox{canonical scaling terms})
\end{eqnarray}
has the ``one-loop'' term (the first line), together with the
``dumbbell'' term (the second line), where the integration with prime
means the integration over shell mode, $1-dt < p < 1$. In the
functional flow equation~\cite{Wetterich:1992yh,Morris:1993qb},
\begin{equation}
 \der_{t} \Gamma_{k} =\half
  \Tr\left[
      \der_{t} R_{k}
      \left(\Gamma_{k}^{(2)}[\phi]+R_{k}\right)^{-1}
      \right],
\end{equation}
where $\Gamma_{k}[\phi]$ is an averaged action with the momentum-shell
parameter $k$, on which the regularization function $R_{k}$ depends,
$\Gamma_{k}^{(2)}$ is the second derivative of $\Gamma_{k}$ with respect
to $\phi$, and $t\equiv \ln(k/\Lambda_{0})$, the whole contributions
come from the ``one-loop'' diagrams. In both cases, the ``one-loop''
diagrams are composed of the full propagators, so that the actual
structure is much more complicated than it appears.

%\comment{nonrelativistic features}

In nonrelativistic systems, because of the absence of antiparticles, the
``one-loop'' diagrams are really one-loop, i.e., the propagators are
bare ones. There are no tadpole-type diagrams, in which an internal line
starts and ends at the same vertex. 

%\comment{two-body sector}

In the two-body sector, there is only one type of one-loop diagrams,
with various kinds of vertices. It gives rise to the RGEs of the
couplings of two-body contact interactions. Note that, since the
two-body amplitude is given as the sum of bubble chains, the cutoff
independence of the two-body amplitude also leads to the same RGEs.

%\comment{apparently only three diagrams contribute}

\begin{figure}
 \includegraphics[width=0.9\linewidth,clip]{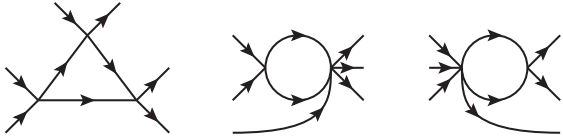}
 \caption{\label{onlythree} Three types of time-ordered one-loop
 diagrams which are apparently the only diagrams contributing to the
 running of three-body coupling constants.}
\end{figure}

For the theory under consideration, if the one-loop property persists in
the three-body sector, there appear to be only three types of one-loop
diagrams that contribute to the running of three-body coupling
constants. See Fig.~\ref{onlythree}. It is clear that the RGE for
$d_{0}$ would not exhibit periodic behavior. To be concrete, the RGE for
$d_{0}$ would be
\begin{equation}
 \frac{dv}{dt} = -4v + \alpha x^3 - \beta vx,
  \label{toosimple}
\end{equation}
where $x$ and $v$ are dimensionless coupling constants defined by
\begin{equation}
 x=\frac{M\Lambda}{4\pi^2}c_{0}, \qquad v=\frac{M\Lambda^4}{6(2\pi^2)^2}d_{0},
\end{equation}
and $t=\ln(\Lambda_0/\Lambda)$ with $\Lambda$ being the floating cutoff,
$\alpha$ and $\beta$ are positive dimensionless constants. In the limit
of infinite scattering length, $x\to -1$, the RGE~\eqref{toosimple} can
be easily solved but the periodic behavior cannot be obtained.  Note
that the inclusion of arbitrary momentum and energy dependent terms
should not alter the conclusion given above because the Efimov effect,
as a universality of long-distance physics, should be incorporated in
the lowest orders in the derivative expansions of the effective field
theory.

It seems that multi-loop diagrams must contribute to the RGE for
$d_{0}$. But how?

\subsection{Normal-ordering}
\label{sec:normal-ordering}

%\comment{operators are usually normal-ordered}

It is important to note that, in nonrelativistic systems, the
interaction operators are all (implicitly) assumed to be normal-ordered.
Under this assumption, we can consider the sectors labeled by the
particle number separately, and the $n$-body operators, such as
$(\phi^\dagger)^n\phi^n$ do not affect the $m$-body sectors with $m<n$.
This is a very favorable feature we would like to keep to simplify the
calculations.

%\comment{operator-formalism implementation}

In order to make the operator structure explicit, let us consider an
operator-formalism implementation of the Wilsonian RG transformation,
where the field operator is Fourier transformed with the magnitude of
momentum $p$ is bounded by the floating cutoff $(0\le p \le
\Lambda)$. The RG transformation amounts to the contractions of only the
shell-modes ($\Lambda-\delta\Lambda < p \le \Lambda$) of the logarithm
of the Dyson operator 
\begin{equation}
 S_{\Lambda-\delta\Lambda}\equiv -i
  \ln
  \left[
  \bra
   T\exp\left[iS_{\Lambda}\right]
  \ket_{\scriptsize\mbox{shell-mode contractions}}
  \right].
\end{equation}
 The field renormalization rescaling is unnecessary in our
nonrelativistic theory. Diagrammatically it is represented as a sum of
one-loop diagrams with the loop momenta being in the shell-mode and with
the operators with lower-momentum modes being attached to the external
lines.

%\comment{counterterms are in general not normal-ordered}

The crucial point is that the counterterms generated by RG
transformations are in general \textit{not} normal-ordered. In order to
obtain the correct RGEs for the coupling constants for the
normal-ordered operators, one needs to rewrite the counterterms in the
normal-ordered form. This rewriting of an $n$-body couterterm leads to
normal-ordered $m$-body operators with $m \le n$.

%\comment{examples}

\begin{figure}
\includegraphics{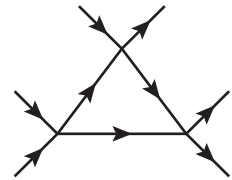} 
\caption{\label{triangle} A time-ordered diagram which contributes to
three-body local counterterms when the loop momentum is set in the
shell. The resulting counterterms are already in the normal-ordered form.}
\end{figure}

\begin{figure}
 \includegraphics{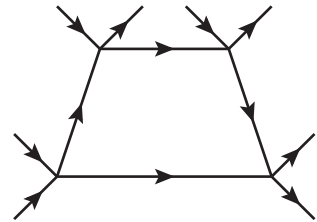}
 \caption{\label{trapezoid} A time-ordered diagram which contributes to
 four-body local counterterms when the loop momentum is set in the
 shell. The resulting counterterms are not normal-ordered. The rewriting
 them in the normal-ordered form generates normal-ordered three-body
 operators as well as normal-ordered four-body operators.}
\end{figure}

Let us explain what is going on by examples. The first example is the
diagram shown in Fig.~\ref{triangle}. With the loop momentum being in
the shell $\Lambda -\delta \Lambda < p \le \Lambda$, it generates an
effective local interaction which can be canceled by $3$-body
counterterms that are all normal-ordered. Note that the diagram is a
time-ordered one. It is impossible to contract any external outgoing and
incoming lines without going backward in time. If we assign an external
outgoing line and an incoming line with the creation and annihilation
operators, $a^\dagger$ and $a$, respectively, this diagram has the
$(a^\dagger)^2(a^\dagger a)a^2$ structure, which is already in the
normal-ordered form. On the other hand, in the second example shown in
Fig.~\ref{trapezoid} with the loop momentum being in the shell, the
diagram generates the contributions that are not normal-ordered. In this
case, it is possible to contract the external lines without going
backward in time. In other words, the diagram has the
$(a^\dagger)^2(a^\dagger a)(a^\dagger a)a^2$ structure which can be
rewritten as a sum of $(a^\dagger)^4a^4$ and $(a^\dagger)^3 a^3$. The
latter contributes to the three-body sector, though naively the diagram
appears to affect only the four-body sector. Note also that a
contraction of the lines amounts to an additional loop, thus the latter
contribution actually comes from the two-loop diagram.

%\comment{Galilean invariant cutoff}

\begin{figure}
 \includegraphics[width=0.5\linewidth,clip]{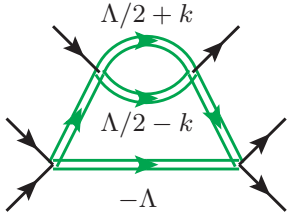}
 \caption{\label{shell-mode} The diagram obtained from
 Fig.~\ref{trapezoid} by normal-ordering. The external energies and
 momenta are set to zero for simplicity. The momenta are assigned so
 that the Galilean invariance is maintained. The double lines indicate
 that the propagators have momenta of order of the cutoff
 $\Lambda$. Note that all the momenta of the internal lines are related
 to the shell-mode momentum.}
\end{figure}

When the diagram with the additional loop is considered, we should treat
it in a Galilean invariant way. We need to impose the cutoff on the
relative momentum of the two lines to maintain Galilean invariance. In
this way, we realize that the ``total momentum'' of the loop must be the
shell mode, while cutoff is imposed on the relative momentum, as shown
in Fig.~\ref{shell-mode}.

%\comment{infinitely many loops}

\begin{figure}
 \includegraphics[width=0.9\linewidth,clip]{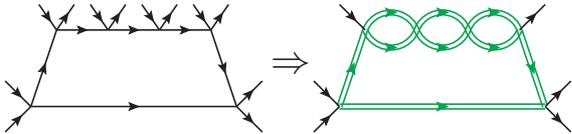}
 \caption{\label{trapezoid2} A time-ordered diagram which contributes to
 six-body local counterterms when the loop momentum is set in the
 shell. The resulting counterterms are not normal-ordered. The rewriting
 them in the normal-ordered form generates normal-ordered three-body
 operators, as well as operators which act only on other sectors. All
 the momenta of the lines of the additional loops are related to the
 shell mode.}
\end{figure}

We go on to the third example in Fig.~\ref{trapezoid2}. The diagram has
six incoming and six outgoing external lines. Naively it only produces
effective six-body operators. However, since it is not normal-ordered,
rewriting it in the normal-ordered form is necessary to obtain the
correct running of the coupling constants for the normal-ordered
operators. There are several ways of contracting the lines. In
particular, it gives rise to three-body operators, which actually come
from four-loop diagrams because the contractions amount to loops. In a
similar way, we can go on to one-loop diagrams with an arbitrary
number of external lines. Again, Galilean invariance forces a particular
momentum assignment so that the momenta of the lines of the additional
loops are related to the shell mode.

%\comment{another kind of diagrams}

\begin{figure}
 \includegraphics[width=0.9\linewidth, clip]{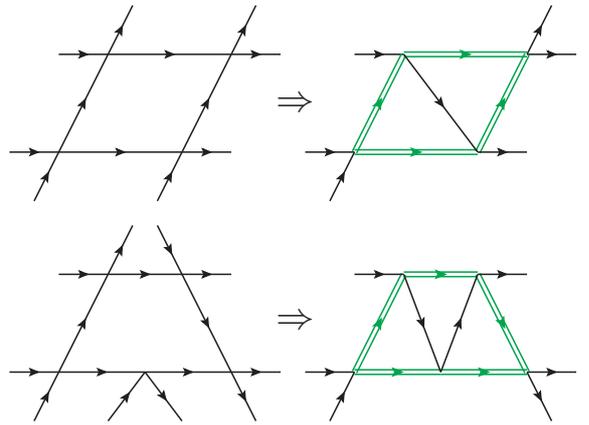}
 \caption{\label{across} Examples of the diagrams in which the
 additional lines get across the shell-mode loop. The momentum
 associated with these lines are not related to the shell mode.}
\end{figure}

There is another kind of diagrams generated by normal-ordering, in which
the contractions give rise to internal lines which get across the
shell-mode loop. See Fig.~\ref{across} for examples. Such diagrams do
not contribute the RGEs. In the usual relativistic field theory, such
diagrams can be obtained after constructing the effective local vertex
by connecting the legs.  That is one of the reasons why the RGEs have
the ``one-loop'' property mentioned above. In the nonrelativistic field
theory, because there is no antiparticle, tadpole-type diagrams cannot
occur. We can just disregard these contractions. The crucial difference
between the diagrams considered in the previous paragraphs and the ones
with lines across the shell-mode loop is that in the former the
additional lines are forced to related to the shell mode, while in the
latter they are not.

%\comment{solution}

The analysis given above provides the solution of the puzzle. The
one-loop property persists even in the three-body sector, but the
generated diagrams are not in general in the normal-ordered form. In
order to obtain the running of the coupling constant for the operators
in the Lagrangian of a nonrelativistic theory, one needs to rewrite the
contributions in the normal-ordered form because the operators in the
Lagrangian are usually assumed to be normal-ordered. In this rewriting,
some contractions may be needed. A contraction amounts to an additional
loop. The diagrams obtained by normal-ordering in which all the lines of
the additional loop are forced to be related to the shell mode, with an
arbitrary number of loops, may contribute to the running of three-body
operators.

%\comment{two-body sector is special}

Note that no multi-loop contribution can occur in the two-body
sector. This is a special feature of the two-body sector. On the other
hand, it is much more involved in many-body sectors. In the three-body
sector, as we will see, even though infinitely many diagrams contribute
to the running of the three-body coupling $d_{0}$, they can be resummed.

\subsection{Breakdown of the naive one-loop property in nonrelativistic
  theory}

In this section, we consider the results obtained in the previous
section from a different side. Let us start with the reason why the
one-loop property arises in the usual formulation of nonperturbative RGEs.

Consider the cutoff dependence of a diagram with an arbitrary number of
loops, whose momenta are cutoff at $\Lambda$. The differentiation with
respect to the cutoff $\Lambda$ picks up the shell mode of each of the
loops, and the result is a sum of the diagrams, in each of which only
one of the loops has the shell-mode momentum. 

Thus, if we have a counterterm which has the same cutoff dependence with
opposite sign as that of the loop in the shell mode, the addition of the
diagram in which each loop of the original diagram is replaced with the
corresponding counterterm cancels the cutoff dependence of the
loop. Note that the diagram with the counterterm has a reduced number of
loops. One can proceed iteratively and determine all the counterterms
which are necessary to make the original diagram finite. This is a
physical picture of how the one-loop property arises.

\begin{figure}
 \includegraphics[width=0.9\linewidth,clip]{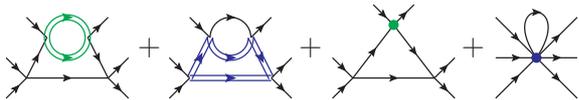}
 \caption{\label{relativistic} The shell-mode diagrams of a two-loop
 example and the corresponding diagrams with the shell-mode loop is
 replaced with the counterterm in the relativistic theory.}
\end{figure}

Let us consider a simple two-loop example shown in
Fig.~\ref{relativistic} to illustrate how the above procedure goes in a
relativistic theory. The loop momenta are cutoff at $\Lambda$. (We have
to work in Euclidean space and impose the condition on the magnitude of
the four-momentum of each propagator in order to maintain Lorentz
(rotational) invariance.) The differentiation of the amplitude with
respect to $\Lambda$ is given by two terms: each has a single loop whose
momentum is in the shell. The point is that we can consider each loop
and the corresponding counterterm individually. The cutoff dependence of
each loop can be compensated by adding the diagram with the
corresponding counterterm. Note that the fourth diagram in
Fig.~\ref{relativistic} contains the tadpole loop. As we
explained in the previous paragraph, this tadpole contribution is
required to make the amplitude cutoff independent.

Let us now suppose the same diagram arises in a nonrelativistic
theory. Apart from the fact that there is no way to maintain Galilean
invariance by imposing a cutoff on the magnitude of the three-momentum
of each propagator, we have a trouble with the tadpole: because of the
absence of anti-particle, there are no tadpole contributions so that we
cannot make the amplitude cutoff independent. This is the reason why the
naive one-loop property must be broken in nonrelativistic theory: the
counterterms obtained by one-loop diagrams do not renormalize the
theory.

A careful examination of the cutoff dependence of multi-loop diagrams in
nonrelativistic theory shows that the loops cannot always be treated
individually because of Galilean invariance. There are contributions in
which more than one loops should be simultaneously in the shell
mode. They compensate the lack of the tadpole contributions absent in
the nonrelativistic theory.  Such contributions are obtained by the
procedure described in the previous section.

\begin{figure}
 \includegraphics[width=0.9\linewidth,clip]{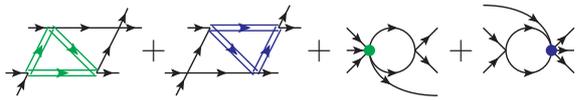}
 \caption{\label{across-shell} The cutoff dependence of the two-loop
 diagram considered in Fig.~\ref{across} can be compensated by the
 counterterms of the individual loops.}
\end{figure}

Note also that the cutoff dependence of the multi-loop diagrams in
Fig.\ref{across} can be compensated by the counterterms of the
individual loops, because the internal lines are not related by the
symmetry so that the loop momenta do not need to be the shell mode
simultaneously. See Fig.~\ref{across-shell}.

\begin{figure}
 \includegraphics[width=0.9\linewidth,clip]{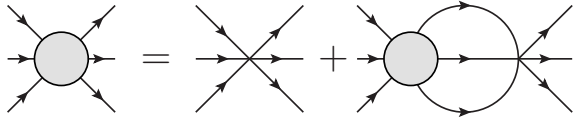}
 \caption{\label{only3body} The Lippmann-Schwinger equation for the
 three-body amplitude in the case where only the three-body interaction
 is present.}
\end{figure}

There is another simple way to see that the naive one-loop property of
RGEs should break down in nonrelativistic theory. Consider the case
where only the $(\phi^{\dagger})^3\phi^3$ interaction is present. In
this case, two-body interactions are not generated (because of the
absence of the tadpole contributions) and the three-body amplitude is
given by a sum of the chains of the two-loop diagrams, depicted in
Fig.~\ref{only3body}. In order to renormalize the amplitude, one needs
the counterterm for the two-loop diagram. The one-loop diagram does not
give rise to such a counterterm because the tadpole cannot contribute in
the nonrelativistic theory. Note that the momenta of the two loops must
be in the shell simultaneously because of the symmetry among the three
propagators and cannot be treated individually.

\section{Derivation of RGEs}
\label{sec:derivation}

%\comment{normal-ordered one-loop contribution}

\begin{figure}[h]
 \includegraphics[width=0.9\linewidth,clip]{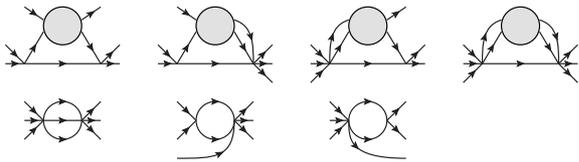}
 \caption{\label{blocks}The NOOL diagrams together with the two genuine
 one-loop diagrams. The shaded blob stands for the two-body amplitude
 defined in Fig.~\ref{two-amp}.}
\end{figure}

\begin{figure}[h]
 \includegraphics[width=0.9\linewidth,clip]{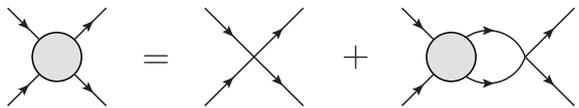}
 \caption{\label{two-amp} The Lippmann-Schwinger equation for the
 two-body scattering amplitude, represented as a shaded blob. The vertex
 actually stands for a collection of vertices, $c_{0}$, $c_{2}$, and
 $b_{2}$ in the present approximation.}
\end{figure}

As we explained in the previous section, there are multi-loop
contributions to three-body operators obtained by normal-ordering
rewriting of the one-loop diagrams that are not in the normal-ordered
form. It turns out that there are five types of such ``normal-ordered
one-loop'' (NOOL) diagrams, together with two genuine (i.e., already
normal-ordered) one-loop ones. They are given in Fig.~\ref{blocks}. Note
that the multi-loop contributions are neatly written in terms of the
two-body scattering amplitude (a blob in Fig.~\ref{blocks}) which
satisfies the Lippmann-Schwinger equation depicted in
Fig.~\ref{two-amp}. Note also that the diagram in Fig.~\ref{triangle}
appears as a part of the first diagram in Fig.~\ref{blocks}.

%\comment{RGEs}

\begin{figure}
 \includegraphics[width=0.9\linewidth,clip]{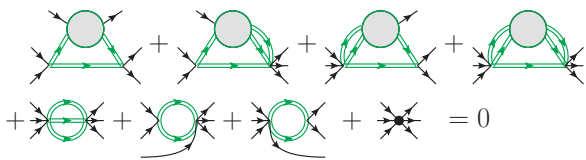}
 \caption{\label{rge3} A diagrammatic representation for the RGEs of the
 three-body operators. Propagators with the momentum being in the shell
 are denoted by double lines.}
\end{figure}

The RGEs can be obtained by setting the loop momenta of the seven
diagrams in the shell and canceling their cutoff dependence by adding
suitable three-body counterterms, as we depicted in Fig.~\ref{rge3}.

It is important to notice that a sum of additional loops, as those in
Fig.~\ref{trapezoid2}, together with the two-body counterterms, forms the
cutoff-independent two-body amplitude so that it does not contribute to
the running of the coupling constants of the three-body operators.
Similarly, the cutoff dependence of the additional loops on the oblique
sides of the trapezoids in the second, third, and the fourth diagrams in
Fig.~\ref{rge3} is already taken into account by the sixth and the
seventh diagrams in Fig.~\ref{rge3}.

%\comment{truncation}

In this section, we consider only the terms that are explicitly shown in
Eq.~\eqref{origEFT}, that is, two-body operators up to including
$\mathcal{O}(p^2)$ and a three-body operator without derivatives. This
truncation makes our RGEs approximate. Higher order corrections are
discussed in Sec.~\ref{sec:corrections}.

%\comment{two-body amplitude}

The two-body amplitude can be treated separately. We assume that the
cutoff $\Lambda$ is small enough so that the two-body amplitude is given
in the effective range expansion form. The renormalized off-shell
amplitude may be written as
\begin{equation}
 {\cal A}(\mu, \bfk_1^{2}, \bfk_{2}^{2})
  \!=\!
  \frac{8\pi}{M}\!
  \left[
   \frac{1}{a_{2}}
   \!-\! \mu 
   \!+\! \frac{1}{2}r_{e}\mu^{2}\!
   \!-\!\frac{3h}{a_{2}}
   \left(
    2\mu^{2}
    \!+\! \bfk_1^{2}
    \!+\! \bfk_{2}^{2}
   \right)
  \right]^{-1}\!\!\!,
  \label{offshell2bodyamp}
\end{equation}
where $a_{2}$ and $r_{e}$ are the scattering length and the effective
range respectively, $\bfk_{1}$ and $\bfk_{2}$ are incoming and outgoing
relative momenta,
\begin{equation}
 \mu=\sqrt{-MP^0 +\bfP^2/4-i\epsilon},
\end{equation}
with $P^\mu=(P^0,\bfP)$ being the total momentum.  (Note that
Eq.~\eqref{offshell2bodyamp} is different by a symmetric factor $1/2$
from that for the ``spinless nucleon'' given in
Ref.~\cite{Harada:2005tw} for which, to mimic the nucleon case, we
intentionally ignore the factor.)  We have an off-shell parameter $h$
coming from the redundant operator. Note that in the on-shell amplitude,
the third term vanishes. The parameters $a_{2}$, $r_{e}$, and $h$ are
cutoff independent.

%\comment{two-body RGEs}

The RGEs for the coupling constants for the two-body operators can be
obtained in Ref.~\cite{Harada:2005tw}:
\begin{eqnarray}
\frac{dX}{dt}
 &=&\left(1-X\right)
 \left(Y+3X^{2}\right)/X,
 \\
\frac{dY}{dt}
 &=&-Y\left(6X^{3}-5X^{2}+2XY-Y\right)/X^{2},
 \\
 \frac{dZ}{dt}
  &=&-\left(6X^{3}Z-3X^{2}Z+2XYZ+Y^{2}\right)/X^{2}\!,
\end{eqnarray}
where $X$, $Y$, and $Z$ are combinations,
\begin{eqnarray}
X&=&1+\frac{1}{3}(y+z),\\
Y&=&x-\frac{1}{5}(y+z)^{2}, \\
Z&=&2y+\frac{1}{3}(y+z)^{2},
\end{eqnarray}
of the dimensionless coupling constants,
\begin{equation}
x=\frac{M\Lambda}{4\pi^{2}}c_{0},\quad
y=\frac{M\Lambda^{3}}{4\pi^{2}}4c_{2},\quad
z=\frac{\Lambda^{3}}{4\pi^{2}}b_{2}.
\end{equation}
The parameter $t$ is defined as $t=\ln(\Lambda_{0}/\Lambda)$, where
$\Lambda_{0}$ is the physical cutoff which is the limit of the
applicability of the EFT, and $\Lambda$ is the floating cutoff.  There
is a nontrivial fixed point, $(X^{*},Y^{*},Z^{*})=(1,-1,-1)$, besides
the trivial one, $(1,0,0)$.

%\comment{the solution of the RGEs}

These RGEs have an analytic solution:
\begin{eqnarray}
X&=&
\frac{(C\Lambda)^{-1}-1}{(C\Lambda)^{-1}-1-C'\Lambda^{2}},
\nonumber \\
Y&=&\frac{(C\Lambda)^{-1}-1}{\left[(C\Lambda)^{-1}-1-C'\Lambda^{2}\right]^{2}},
\nonumber \\
Z&=&\frac{C''\Lambda-1}{\left[(C\Lambda)^{-1}-1-C'\Lambda^{2}\right]^{2}},
\label{RGEs2}
\end{eqnarray}
where $C$, $C'$, and $C''$ are the integration constants. 
They are related to the effective range expansion parameters, $a_{2}$,
$r_{e}$, and $h$, by
\begin{equation}
C=\frac{2a_{2}}{\pi},\quad
C'=h, \quad
C''=\frac{\pi}{4}r_{e}.
\label{integration_consts}
\end{equation}

%\comment{three-body RGEs}

In terms of them, the three-body RGEs are written as follows:
\begin{eqnarray}
\frac{dv}{dt}&=&
 \left[
  3b^{2}\frac{1}{S^{2}}
  \left(
   T-\frac{2}{b}U
  \right)
  -c
 \right]
 v^{2}
 \nonumber \\
 &&{}
  +
  \left[
   12b\frac{V}{S^{2}}
   \left(
    T-
    \left(
     3+\frac{1}{b}
    \right)
    U
   \right)
   -4-6V
  \right]
  v
  \nonumber \\
 &&{}
  +12
  \frac{V^{2}}{S^{2}}
  \left(
   T-6U
  \right) 
  \label{RGEs3}
\end{eqnarray}
where we have introduced several combinations,
\begin{eqnarray}
 S&=&X^{2}-aY, \\
 T&=&YS+\frac{3}{4}(X^{2}Z+Y^{2}), \\
 U&=& X\left(X-1\right)\left(X^{2}+Y\right), \\
 V&=&Y+3(X-1)+\frac{9}{5}(X-1)^{2},
\label{combinations}
\end{eqnarray}
together with numerical constants,
\begin{eqnarray}
 a&=&
  \frac{\sqrt{3}}{4}\pi-1,
  \nonumber \\
 b&=&
  1-\frac{\sqrt{3}}{4}\pi
  +\frac{\sqrt{3}}{2}
  \arctan\frac{\sqrt{3}}{2},
  \nonumber \\
 c&=&
  \frac{7}{3}
  -\frac{\sqrt{3}}{4}\pi
  -\frac{7\sqrt{3}}{18}
  \arctan\frac{\sqrt{3}}{2}.
  \label{bc}
\end{eqnarray}
Note that in these equations, $X$, $Y$, and $Z$ are the solution,
Eq.~\eqref{RGEs2}, with the integration constants,
Eq.~\eqref{integration_consts}. 

In Appendix~\ref{sec:appendix:NOOL}, a sample calculation of a
contribution to the RGE is given for the purpose of illustration.
It is a bit difficult to find out the shell mode contribution of
the two-loop diagram shown in the fifth diagram in Fig.~\ref{rge3}. In
Appendix~\ref{sec:appendix:twoloop}, we explain how to identify the
shell mode contribution.

%\comment{$v^2$ term appears}

It is important to note that the RGE now has terms proportional to
$v^2$. The appearance of these terms allow the periodic behavior of $v$
in the RG evolution, as we will see in the next section.

%\comment{building blocks of the three-body amplitude}

It is interesting to note that these seven diagrams in Fig.~\ref{blocks}
are the irreducible building blocks of which the three-body amplitude is
composed, just as simple two-point bubbles are the building blocks of
the two-body amplitude.

%\comment{comparison with dimeron formulation}

\begin{figure}
 \includegraphics[width=0.9\linewidth,clip]{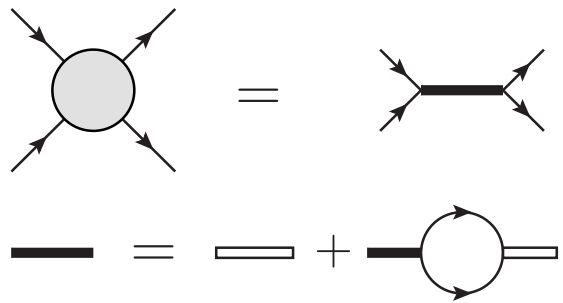}
 \caption{\label{identity} The dimeron representation of the two-body
 amplitude, composed of the dressed dimeron propagator and the
 two-particle-dimeron vertices at the ends.}
\end{figure}

The amplitude composed of these blocks can be compared with that
obtained in the formulation with the dimeron field. The two-body
amplitude is equal to the (dressed) dimeron propagator multiplied by the
factor $g_{1}^2$ coming from the two-particle-dimeron vertices at the
ends. See Fig.~\ref{identity}. With this identification, one easily sees the
correspondence between the three-body amplitude in our formulation and
that in the formulation with dimeron.

%\comment{difference}

There is however a difference: in our formulation, a one-loop diagram
with a two-body vertex and a three-body vertex (which we call a
``two-three loop'') contributes, while in the existing calculations with
dimeron these contributions are not included.

%\comment{how the two-three loop affects}

Note that the cutoff dependence of the two-three loop cannot be
renormalized as a part of two-body amplitude because the two-body
amplitude is composed solely of two-body vertices, but is canceled by
three-body counterterms, as we explained earlier. See the sixth and the
seventh diagrams in Fig.~\ref{rge3}.  The momentum dependence becomes
important when the two-three loop is embedded in the three-body
amplitude. When the shell-mode momentum is considered as in the second,
third, and the fourth diagrams in Fig.~\ref{rge3}, the momentum
dependence of the two-three loops becomes an additional source of the
cutoff dependence.

%\comment{$D\phi$-$D\phi$ vertex}

\begin{figure}
 \includegraphics[width=0.9\linewidth,clip]{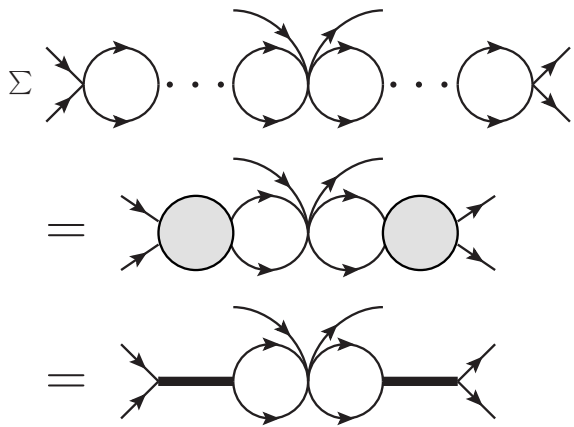}
 \caption{\label{DphiDphi}The insertion of the three-body vertex in the
 two-body amplitude. In the third line, the two-body amplitudes are
 replaced by the full dimeron propagator. Because of the two-three
 loops, it is not equal to the (momentum-independent) $D\phi$-$D\phi$
 vertex.}
\end{figure}

In the existing calculations with dimeron, the $D\phi$-$D\phi$ vertex is
momentum independent even for the full dimeron lines. On the other hand,
as depicted in Fig.~\ref{DphiDphi}, a single insertion of the three-body
vertex in the two-body amplitude cannot be represented by a
momentum-independent $D\phi$-$D\phi$ vertex, but has momentum dependence
due to the two-three loops.

\section{RG flows and the Efimov parameter}
\label{sec:RGflow}

\subsection{Flows in the $r_e=h=0$ subspace}

%\comment{$\xi$-v flow}

\begin{figure}
 \includegraphics[width=0.9\linewidth,clip]{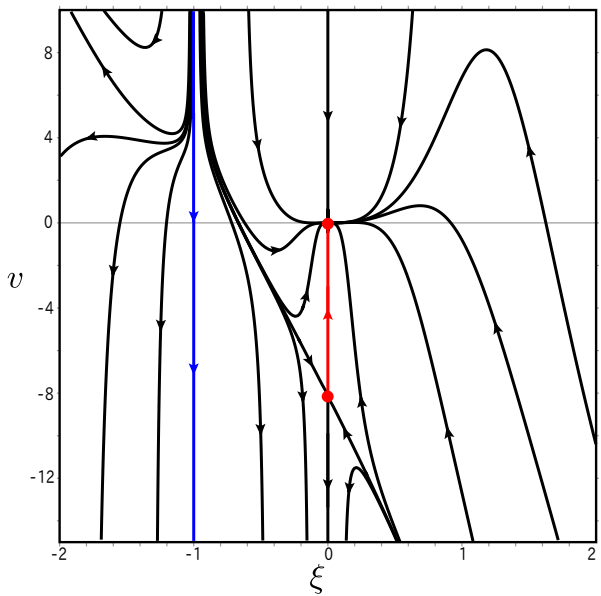}
 \caption{\label{xv}The RG flow in the $\xi$-$v$ plane. The
 arrows indicate the directions of the flow to the infrared (larger
 $t$). }
\end{figure}

Let us begin, for simplicity, with the RG flows on which the effective
range, $r_{e}$, and the off-shell parameter, $h$, are zero. In this
case, the RGEs are drastically simplified and are written in terms of
$\xi\equiv Y/X^2$, as
\begin{eqnarray}
 \frac{d}{dt}\xi &=& -\xi(1+\xi), 
  \label{RGExi}\\
 \frac{d}{dt}v &=& \frac{12\xi^3}{1-a\xi} 
  +\left(
    \frac{12b\xi^2}{1-a\xi}-4-6\xi
   \right)v
  \nonumber \\
 &&{}
  +\left(
    \frac{3b^2\xi}{1-a\xi}-c
   \right)v^2.
  \label{dvdt}
\end{eqnarray}
Note that the RGE for $\xi$ is the same as that for $x$ in the leading
order approximation. The RG flow in the $\xi$-$v$ plane are shown in
Fig.~\ref{xv}. The line $\xi=-1$ corresponds to the nontrivial fixed
point of the two-body RGEs. The flow there goes down periodically,
exhibiting the limit cycle behavior.  In addition to the trivial fixed
point, we find a nontrivial fixed point $(\xi_\star,
v_\star)=(0,-4/c)=(0,-8.126\cdots)$, which we call the \textit{Borromean
fixed point}, though it apparently has nothing to do with the known
Borromean systems.  In the full set of RGEs, the Borromean fixed point
is at $(x_{\star}, y_{\star}, z_{\star}, v_{\star})=(0,0,0,-4/c)$, so
that there are no two-body interactions but only the three-body
interaction.

%\comment{pole at $\xi=1/a$}

One may notice that there is a singularity in Eq.~\eqref{dvdt} at
$\xi=1/a$. This is due to the pole of the two-body amplitude which
corresponds to the bound state. See Appendix~\ref{sec:appendix:pole} for
more details.

\subsection{Limit cycle}

%\comment{limit cycle}

The limit cycle behavior is due to the $v^2$ term in the RGE for
$v$. With $(x,y,z)=(-1,-1/2,1/2)$ ($(X,Y,Z)=(1,-1,-1)$), where,
$a_{2}=\infty$ and $r_{e}=h=0$, the RGE for $v$ becomes
\begin{equation}
 \frac{dv}{dt}=A+Bv+C v^{2},
\end{equation}
with
\begin{equation}
 A=-\frac{48}{\sqrt{3}\pi},\ 
 B=2+\frac{48b}{\sqrt{3}\pi},\ 
 C=-\frac{12b^{2}}{\sqrt{3}\pi}-c,
\end{equation}
so that the discriminant is negative,
\begin{equation}
 D\equiv B^{2}-4AC=-4.27374\cdots <0.
\end{equation}
Thus the solution is given by
\begin{eqnarray}
 v=\frac{-B+\sqrt{-D}\tan
  \left(
   \frac{\sqrt{-D}}{2}(t+t_{0})
  \right)}{2A},
  \label{sol-v}
\end{eqnarray}
where $t_{0}$ is a constant of integration. It shows the periodic
behavior with the period
\begin{eqnarray}
T=\frac{2\pi}{\sqrt{-D}}\simeq 3.03932.
\end{eqnarray}
In terms of the Efimov parameter, $s_{0}$, it is written as
\begin{eqnarray}
 s_{0}=\frac{\pi}{T}\simeq 1.03365,
\end{eqnarray}
which is only $2.7\%$ off the Efimov's value.

%\comment{difference is due to the 2-3 loop}

It is interesting to compare the value with that obtained in the
``point-like'' approximation in the dimeron
formulation~\cite{Moroz:2008fy}, $s_{0}\simeq 1.393$, which is about
$40\%$ off the Efimov's value. The difference is due to the
contributions from the two-three loops, as we explained in the previous
section.

\subsection{Off-critical cases}

%\comment{off critical}

\begin{figure}
 \includegraphics[width=\linewidth,clip]{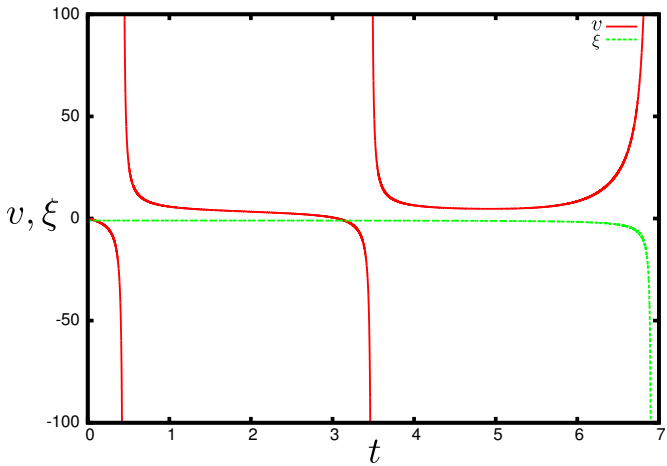}
 \caption{\label{2b-strong}The running of $v$ and $\xi$ as functions of
 $t$ in the two-body strong-coupling phase ($\xi<-1$). The initial value
 is taken as $v(t=0)=0$ and $\xi(t=0)=-1.001$.}
\end{figure}

\begin{figure}
 \includegraphics[width=\linewidth,clip]{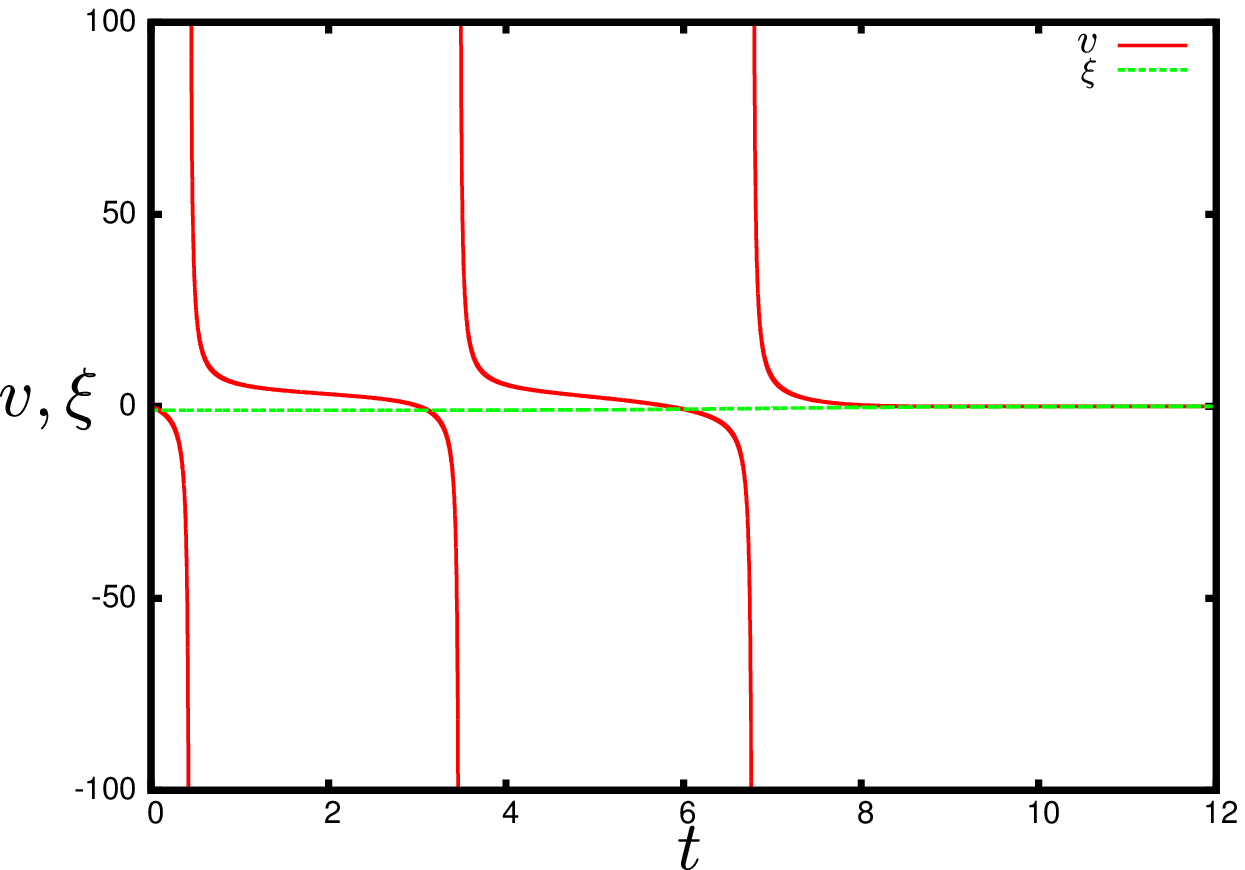}
 \caption{\label{2b-weak}The running of $v$ and $\xi$ as functions of
 $t$ in the two-body weak-coupling phase ($\xi>-1$). The initial value
 is takes as $v(t=0)=0$ and $\xi(t=0)=-0.999$.}
\end{figure}

We can investigate the off-critical, i.e., finite scattering length,
behavior. Fig.~\ref{2b-strong} and Fig.~\ref{2b-weak} show the running
of $v$ and $\xi$ as functions of $t$ starting with $\xi(t=0)+1=-0.001$
and $\xi(t=0)+1=0.001$, respectively. If the flow is close enough to the
critical line ($\xi=-1$), $v$ diverges (to negative infinity) a finite
number of times before going to the trivial fixed point (for $\xi>-1$)
or to the positive infinity (for $\xi<-1$). We suspect that the
occurrence of divergences corresponds to the existence of three-body
bound states, as for the critical case in which the flow diverges
infinite times, corresponding to an infinite number of bound
states. Thus, if the flow is close enough to the critical line, there
are a finite number of three-body bound states.


%\comment{two-body v.s. three-body}

The behavior in the two-body strong-coupling phase ($\xi<-1$) is
particularly interesting from the NEFT point of view. In the
$^3S_1$-$^3D_1$ channel, the two-nucleon system is in the
strong-coupling phase and very close to the critical case. There is a
shallow two-body bound state, deuteron. The binding energy of the triton
($E_{t}=8.48$ MeV) which is in the spin one-half channel in the
neutron-deuteron system, is larger than that of deuteron ($E_{d}=2.2246$
MeV). Even though the present theory is bosonic, we expect that the same
RG structure persists in the fermionic theory, and that the interplay
between the three-body and the two-body bound states of the bosonic
theory may explain that of the nucleon systems.

%\comment{divergences indicates the existence of bound states}

The coupling $\xi$ diverges to negative infinity at a finite value of
$t$ ($t_2$) in the two-body strong-coupling phase. It implies the
existence of the two-body bound state and the value $t_2$ would
correspond to the binding energy of it. Once $\xi$ diverges to negative
infinity, the flow appears from the positive infinity of $\xi$,
decreasing to the trivial fixed point, $\xi=0$. (The continuity may be
easily seen by introducing a change of variable from $\xi$ to
$\theta\equiv \tan^{-1}\xi$. The RGE~\eqref{RGExi} becomes $d\theta/dt =
-\sin\theta(\cos\theta+\sin\theta)$. The transition from the negative
infinity to the positive infinity of $\xi$ corresponds to passing
$-\pi/2$ of $\theta$.)

%\comment{binding energies}

When $v$ diverges to negative infinity in the two-body strong-coupling
phase ($\xi<-1$), it occurs before $\xi$ diverges. In the positive $\xi$
region, on the other hand, we numerically find, by examining the flow,
that all the flows which come from the two-body strong-coupling phase never
diverge to negative infinity. The above observation implies that, if
both the two-body and the three-body bound states exist, the binding
energy of the two-body bound state is always smaller than the three-body
bound state.

\subsection{Borromean fixed point}
\label{sec:borromean}

%\comment{Borromean fixed point}

To our best knowledge, the existence of the Borromean fixed point has
never been noticed in the literature. The existence of it and of the
critical surface on which the Borromean fixed point resides implies a
new phase where the two-body interactions do not support the two-body
bound states, but the strong three-body interaction gives rise to the
three-body bound states.

%\comment{linearized RGE}

We linearize the RGE near the fixed point, $(\xi,v)=(\xi_{\star},v_{\star})
+ (\delta \xi, \delta v)$,
\begin{eqnarray}
 \frac{d}{dt}
  \left(
   \begin{array}{c}
    \delta \xi\\
    \delta v
   \end{array}
  \right)
 &=& 
 \left(
  \begin{array}{cc}
   -1 & 0\\
   \frac{24}{c}\left(1+\frac{2b^2}{c}\right) & 4
  \end{array}
 \right)
 \left(
  \begin{array}{c}
   \delta \xi\\
   \delta v
  \end{array}
 \right).
\end{eqnarray}
By diagonalizing it, we see that the scaling dimension of the relevant
coupling is $+4$. In the next section, we investigate the Borromean
fixed point once again with higher order corrections.

\subsection{Effects of nonzero values of $r_e$ and $h$}

%\comment{h- and r-dependence}

\begin{figure}
 \includegraphics[width=0.9\linewidth,clip]{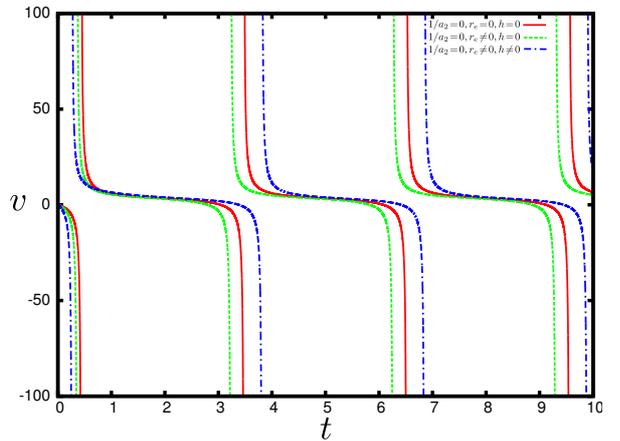}
 \caption{\label{hidden} The effects of nonzero values of $h$ and
 $r_e$. The red line is the solution Eq.~\eqref{sol-v}, with
 $r_{e}=h=0$. The green line is the case with $r_{e}\Lambda_{0}=1$ and
 $h=0$, where $\Lambda_{0}$ is the physical cutoff. The blue line is the
 case with $r_{e}\Lambda_{0}=h\Lambda_{0}^2=1$. Three of them have
 asymptotically the same period.}
\end{figure}

Going back to the full set of RGEs, Eq.~\eqref{RGEs3} together with
Eqs.~\eqref{RGEs2}, we can investigate the effects of nonzero values of
$h$ and $r_e$. The results are shown in Fig.~\ref{hidden}. It is
interesting to note that they affect the flow of $d_{0}$, especially the
``period,'' even though their effects damp as $t$ grows. Physically
speaking, it implies that the ratios of the binding energies of the
three-body bound states varies and asymptotically becomes the universal
value $e^{-2\pi/s_{0}}$ for the bound states accumulate toward the
threshold. It is surprising to see that the parameter, $h$,
corresponding to the redundant operator has such observable effects,
even though it does not contribute to the two-body
observables. Furthermore, it does not even appear in the off-shell
amplitude, Eq.~\eqref{offshell2bodyamp}, in the critical case
($1/a_{2}=0$)!

\section{Higher order corrections}
\label{sec:corrections}

In this section, we concentrate on the cases in which the two-body
sector is on the nontrivial and the trivial fixed points, and consider
the effects of higher order three-body operators. Note that all the
irrelevant two-body operators do not contribute in these cases.

%\comment{added operators}

We include the following three-body operators,
\begin{eqnarray}
 {\cal L}_{\mbox{\scriptsize higher}}\!\!&=&\!\!
  \frac{d_{2}}{48}
  \left[
   \left(
    \phi^{3}
   \right)^{\dagger}\!
   \phi
   \left(
    \phi
    \overleftrightarrow{\nabla}^{2}
    \phi
   \right)
   \!+\!{\mbox h.c.}
  \right]
  \nonumber \\
 & &{}
  +
  \frac{e_{2}}{24}
  \left[
   \left(
    \phi^{3}
   \right)^{\dagger}\!
   \phi
   \left\{
    \phi\left(
         i\der_{t}+\frac{\nabla^{2}}{2M}
        \right)\phi
        \right\}
   \!+\!{\mbox h.c.}
  \right],
  \nonumber \\
\end{eqnarray}
and obtain the RGEs for the dimensionless coupling
constants, $v$, $s$, and $w$, the latter two of them are defined as
\begin{eqnarray}
% v=\frac
% {M\Lambda^{4}}
% {6(2\pi^{2})^{2}}
% d_{0},
% \quad
 u=\frac{M\Lambda^{6}}{(2\pi^{2})^{2}}d_{2},\quad
 w=\frac{\Lambda^{6}}{(2\pi^{2})^{2}}e_{2},\quad
 s=u+\frac{1}{3}w.
\end{eqnarray}

\subsection{On the nontrivial fixed point}

%\comment{RGEs}

On the nontrivial fixed point of the two-body sector,
$(x,y,z)=(-1,-1/2,1/2)$, RGEs are given by
\begin{widetext}
\begin{eqnarray}
 \frac{dv}{dt} \!&=&\!
  -\frac{48}{\sqrt{3}\pi}
  +\left(
    2+\frac{48b}{\sqrt{3}\pi}
   \right)v
  -\left(
    \frac{12b^{2}}{\sqrt{3}\pi}+c
   \right)v^{2}
  +\left(
    \frac{3}{2}+\frac{4}{\sqrt{3}\pi}
   \right)s
  -\left(
    \frac{7}{120}+\frac{1}{12\sqrt{3}\pi}
   \right)s^{2}
  -\left(
    \frac{2b}{\sqrt{3}\pi}+\frac{1}{3}
   \right)vs,\label{higherv}
  \\
%%%%%%%%%%%%%%
 \frac{ds}{dt} \!&=&\!
 \frac{64}{3\sqrt{3}\pi}
 -\left(
   2-\frac{24b}{\sqrt{3}\pi}+\frac{8}{9\sqrt{3}\pi}
  \right)s
 -\left(
   \frac{1}{6}+\frac{b}{\sqrt{3}\pi}
  \right)s^{2}
 -\left(
   2+\frac{32b}{3\sqrt{3}\pi}
  \right)v
 -\left(
   \frac{12b^{2}}{\sqrt{3}\pi}+c
  \right)vs,\label{highers}
\\
% \nonumber \\
% \end{eqnarray}
% \begin{eqnarray}
% %%%%%%%%%%%%%%%%%%
 \frac{dw}{dt}\!&=&\!
 \frac{576}{\sqrt{3}\pi}
 +\frac{48b}{\sqrt{3}\pi}w
 -24\left[
     1+\frac{8}{\sqrt{3}\pi}
     \left(
      b+\frac{4}{7}
     \right)
    \right]v
 -\left[
   6+\frac{8}{\sqrt{3}\pi}
   \left(
    9b+4
   \right)
  \right]s
 +6\left[
    e+\frac{8b}{\sqrt{3}\pi}
    \left(
     b+3d
    \right)
   \right]v^{2}
 \nonumber \\
 &&{}
  +\left[
    \frac{1}{4}+\frac{4}{\sqrt{3}\pi}
    \left(
     \frac{3}{4}b+\frac{1}{12}
    \right)
   \right]s^{2}
  +\left[
    3c+\frac{4}{\sqrt{3}\pi}
    \left(
     9b^{2}+2b+3d
    \right)
   \right]vs
  -2\left(
     \frac{12b^{2}}{\sqrt{3}\pi}+c
    \right)vw
  -\left(
    \frac{2b}{\sqrt{3}\pi}+\frac{1}{3}
   \right)sw,
  \nonumber \\
\end{eqnarray}
\end{widetext}
where the constants $b$, $c$ are given in Eqs.~\eqref{bc} and $d$ and $e$
are defined as 
\begin{eqnarray}
 d
% &=&
% \frac{\sqrt{3}}{3}\arctan
% \frac{2}{\sqrt{3}}
% -\frac{2}{7}
% \nonumber \\
  &=&
   -\frac{2}{7}
   +\frac{\sqrt{3}}{6}\pi
   -\frac{\sqrt{3}}{3}\arctan
   \frac{\sqrt{3}}{2},
\nonumber \\
 e&=&
  -\frac{2}{3}
  +\frac{\sqrt{3}}{6}\pi
  +\frac{\sqrt{3}}{9}\arctan
  \frac{\sqrt{3}}{2}.
\end{eqnarray}
Note that RGEs for $v$ and $s$ do not depend on $w$, so that they can be
solved without solving the RGE for $w$. In the following we mainly
consider the flows in the $v$-$s$ plane.

\begin{figure}
 \includegraphics[width=0.9\linewidth,clip]{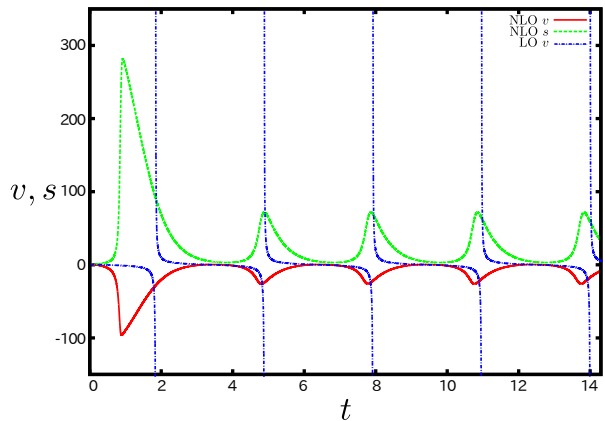}
 \caption{\label{limitcycle} The $t$ dependence of the three-body
 coupling constants, $v$ and $s$. The blue line stands for the leading
 order solution of $v$. The initial values are $(v_{0},s_{0})=(0.9,
 0.7)$.}
\end{figure}

\begin{figure}
 \includegraphics[width=0.9\linewidth,clip]{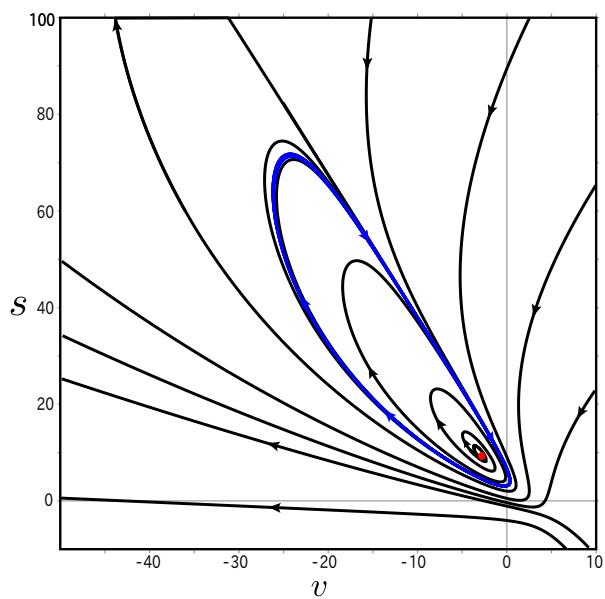}
 \caption{\label{lc-loop} The limit cycle behavior in the $v$-$s$
 plane. The blue loop is the limit cycle. Inside the loop there is a
 nontrivial fixed point (red bullet).}
\end{figure}

These RGEs can be solved numerically. The $t$ dependence of the coupling
constants, $v$, $s$, together with the leading order one for $v$, is
shown in Fig.~\ref{limitcycle}. Note that after an initial transient
region, they exhibit periodic behavior. Note also that they do not
diverge in the periodic region. The RG flow in the $v$-$s$ plane is
shown in Fig.~\ref{lc-loop}.

%\comment{Efimov parameter}

We numerically read off the period of the limit cycle as $2.9810$, which
corresponds to the value of Efimov parameter $s\simeq 1.0539$. The value
is about $4.73\%$ off the Efimov's value. Compared with the leading
order result, it deviates a bit more from the Efimov's value, though it
is still very close. It might suggest that the convergence is not
monotonous.

%\comment{different from the leading order result-1}

It is interesting to see that the limit cycle behavior emerges very
differently from the leading-order one discussed in the previous
section. In the leading order, since there is only one coupling, the
periodicity is not possible unless it gets through the infinity. On the
other hand, in the higher order, it is possible for the flows to have a
periodicity while staying finite. It explains why the drastic change of
the behavior of the RG flows occurs when the higher order contributions
are taken into account.

%\comment{difference from the leading order result-2}

\begin{figure}
 \includegraphics[width=0.9\linewidth,clip]{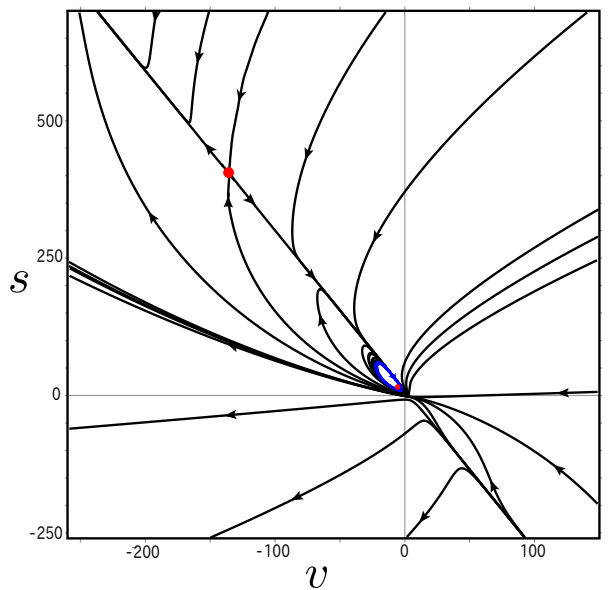}
 \caption{\label{wide3body} The RG flows in the $v$-$s$ plane in a wide
 view. Another fixed point is shown together with the limit cycle loop.}
\end{figure}

Another difference comes from the existence of the transient region. In
the leading order, exact periodicity appears from the
beginning. Actually the structure of the RG flow in the higher order is
much richer. First of all, there is a nontrivial fixed point,
numerically found to be at $(v_\star, s_\star, w_\star)=(-3.14, 10.05,
-60.73)$. As shown in Fig.~\ref{lc-loop}, the flows from this nontrivial
fixed point gradually approach to the limit cycle. On the other hand,
the flows outside the limit cycle loop are grouped in the two
categories: some flows are directly attracted by the limit cycle and
gradually approach to it, and the others go to infinity (and come back
from the opposite side) and then approach to the limit cycle. The
existence of these two categories is related to another nontrivial fixed
point at $(-135.52, 403.62, -1069.65)$. See Fig.~\ref{wide3body}.

%\comment{eigenvalues}

The eigenvalues of the linearized RGEs at the fixed point located inside
the limit-cycle loop are numerically obtained as $0.3288\pm 2.3212i$ and
$1.9817$. The eigenvector of the real positive eigenvalue is in the $w$
direction. It is known that the imaginary part of the complex
eigenvalues, $\theta$, gives approximately the period of the limit
cycle,
\begin{equation}
 T\simeq \frac{2\pi}{\mbox{Im}\; \theta}.
\end{equation}
With $\theta=2.3212$, we find $T=2.7069$, which corresponds to the value
of Efimov parameter $s_{0}=1.1606$.

\subsection{On the trivial fixed point}

As explained in Sec.~\ref{sec:borromean}, we find a nontrivial fixed
point of the three-body coupling $d_{0}$ on the two-body trivial fixed
point. In this section, we examine the effects of the higher order
operators on it.

%\comment{RGEs}

The RGEs on the trivial fixed point of the two-body sector are given by
\begin{eqnarray}
 \frac{dv}{dt}
  \!&=&\!
  -4v-c v^{2}-\frac{1}{3}vs-\frac{7}{120}s^{2},
  \nonumber \\
 \frac{ds}{dt}
  \!&=&\!
  -6s-cvs-\frac{1}{6}s^{2},
  \nonumber \\
 \frac{dw}{dt}
  \!&=&\!
  -6w+6ev^{2}\!+\!\frac{1}{4}s^{2}\!
%   \nonumber \\
%  &&{}
  +3cvs-2cvw-\!\frac{1}{3}sw.
\end{eqnarray}

%\comment{RG flow in the v-s plane}

\begin{figure}
 \includegraphics[width=0.9\linewidth,clip]{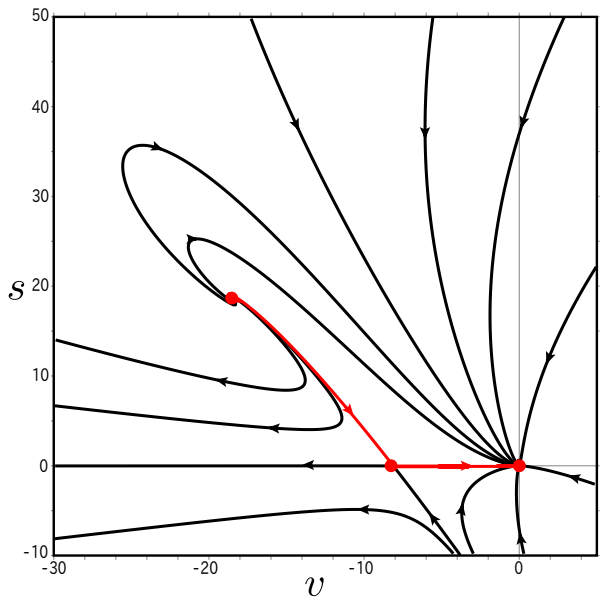}
 \caption{\label{bfp} The RG flow on the two-body trivial fixed point
 ($\xi=0$). We find another nontrivial fixed point in addition to the
 trivial and Borromean fixed points.}
\end{figure}

The RG flow in the $v$-$s$ plane is shown in Fig.~\ref{bfp}. We find
the Borromean fixed point at the same value of $v$. In the $v$-$s$-$w$
space, the fixed point is at
$(-4/c,0,-48e/c^2)\approx(-8.13,0,-74.80)$. This strongly suggests that
the existence of the Borromean fixed point is not an artefact of the
restriction of the set of operators.

%\comment{scaling dimension at the Borromean fixed point}

By linearizing the RGEs at the Borromean fixed point, we find the
eigenvalues, $\nu$, and the corresponding (unnormalized) eigenvectors,
$(\delta v, \delta s, \delta w)$;
\begin{eqnarray}
 4: \left(
         \begin{array}{c}
          c \\
          0 \\
          24e\\
         \end{array}\right),\quad
 2: \left(
         \begin{array}{c}
          0 \\
          0 \\
          1\\
         \end{array}\right),\quad
 -2: \left(
         \begin{array}{c}
          2c/3 \\
          -3c^2 \\
          4e-9c^2\\
         \end{array}\right).
\end{eqnarray}
Note that there are two relevant operators. 

%{\color{red}There must be an operator of dimension 0}

%\comment{other fixed points}

We also find other fixed points at $(-18.45,18.48,-58.81)$ and at
$(-247.09,693.75,-917.51)$. The former is located at the center of the
whirl, and the latter, which is not displayed in Fig.~\ref{bfp}, is
related to the fixed point shown in Fig.~\ref{wide3body}.

%\comment{bifurcation}

\begin{figure}
 \includegraphics[width=0.9\linewidth,clip]{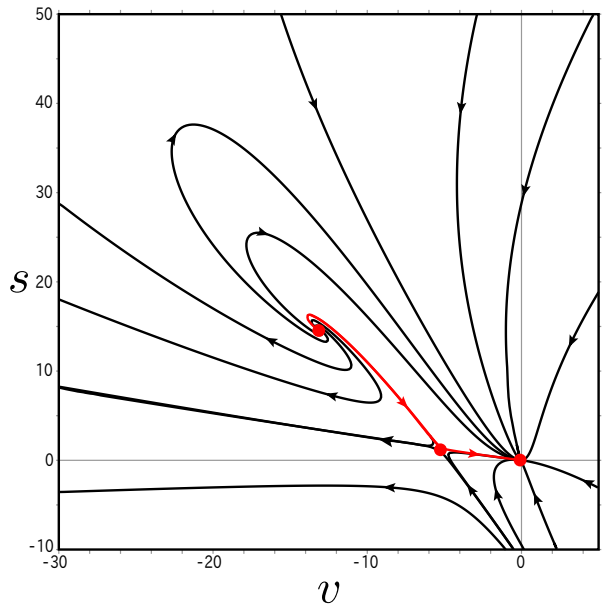}
 \caption{\label{xi-0.25} The RG flow in the $v$-$s$ plane with
 $\xi=-0.25$. The Borromean fixed point and the trivial fixed point move
 together and eventually fuse and disappear.}
\end{figure}

\begin{figure}
 \includegraphics[width=0.9\linewidth,clip]{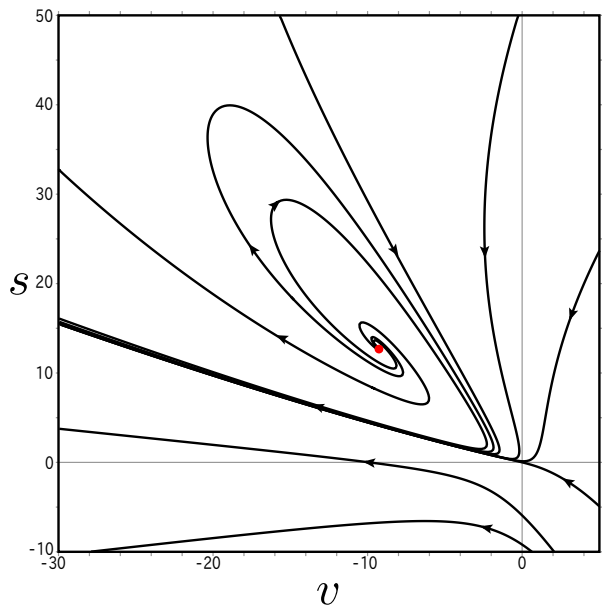}
 \caption{\label{xi-0.5} The RG flow in the $v$-$s$ plane with
 $\xi=-0.5$. There is no Borromean nor trivial fixed point.}
\end{figure}

\begin{figure}
 \includegraphics[width=0.9\linewidth,clip]{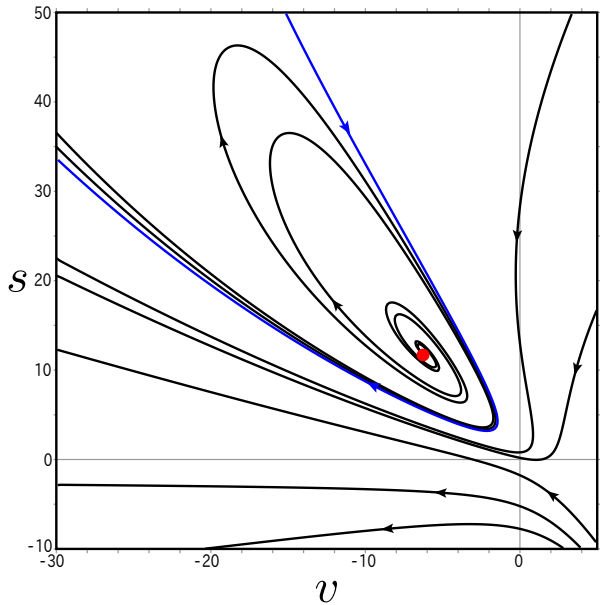}
 \caption{\label{xi-0.75} The RG flow in the $v$-$s$ plane with
 $\xi=-0.75$. A large limit cycle loop emerges.}
\end{figure}

Actually Fig.~\ref{bfp} transforms into Fig.~\ref{wide3body} (or
Fig.~\ref{lc-loop}) if $\xi$ is treated as a parameter changing from
zero (the trivial fixed point) to $-1$ (the nontrivial fixed point
approached from the weak-coupling phase). See Fig.~\ref{xi-0.25} to
Fig.~\ref{xi-0.75}. The Borromean fixed point and the fixed point at the
origin move together to fuse and disappear, while the whirl becomes
enclosed by the limit cycle. This is an example of bifurcation.


\section{Summary}
\label{sec:summary}

%\comment{what we did in this paper}

In this paper, we performed the Wilsonian RG analysis of nonrelativistic
three-body systems of identical bosons without introducing the dimeron
field. We explained why the multi-loop diagrams contribute to the RGEs
and obtained the RGEs in the restricted space of operators. The periodic
behavior in the case of infinite two-body scattering length is obtained
with the value of the Efimov parameter being very close to the Efimov's
value. The difference between our result and that obtained by using the
dimeron field is clarified.

%\comment{rich structure}

We found very rich structures of the RG flows: beside the periodic
behavior responsible for the Efimov effect, we found the Borromean fixed
point, which, to our best knowledge, has never been noticed in the
literature. It implies the existence of the three-body strong-coupling
phase. We also investigated the effects of finite values of the
effective range and off-shell parameters, and found that they affects
the ratios of the binding energies of the three-body bound states.

%\comment{triton}

We also try to explain that the binding energy of triton is larger than
that of deuteron on the basis of the study of the RG flows in our
bosonic system.

%\comment{by extending the space of three-body operators}

We then extended our set of three-body operators to $\mathcal{O}(p^2)$
and considered the case of the two-body fixed points. On the two-body
nontrivial fixed point, the limit cycle of the three-body couplings is
realized as a loop of a finite size, and two nontrivial fixed points are
found. On the two-body trivial fixed point there is a nontrivial fixed
point around which a whirling of the RG flow occurs, in addition to the
Borromean and the trivial fixed points.

%\comment{three-body strong coupling phase}

The three-body strong-coupling phase, with the phase boundary on which
the Borromean fixed point resides, has a clear physical picture: it
corresponds to the situation in which there are no (or very weak)
two-body interactions, but a strong (short-range) attractive three-body
interaction among the particles. We do not know if such a situation is
possible in the real world.

\begin{acknowledgments}
 I.~Y. is grateful to H.~Makino for his help in the early state of the
 investigation. The authors would like to thank J.~Gegelia for the
 discussions and the comments on the first version of the manuscript.
 This work was supported by JSPS KAKENHI (Grant-in-Aid for Scientific
 Research (C)) (22540286).
\end{acknowledgments}

\appendix

\section{Use of a dimeron}
\label{sec:appendix:dimeron}

In this appendix, we illustrate a way of introducing a dimeron field in
the path integral formalism, and emphasize that it is difficult to
obtain the relations between the coupling constants of the original
theory and those in a theory with a dimeron.

Let us start with the (Euclidean) partition function,
\begin{equation}
 Z=\int d\phi^\dagger d\phi e^{-\int d^4x \;\mathcal{L}_{E}},
\end{equation}
where $\mathcal{L}_{E}$ is the Euclidean version of the Lagrangian given
by Eq.~\eqref{origEFT}. We insert the identity,
\begin{eqnarray}
1=\int dD^{\dagger}dD
\,\,e^{-\int d^4x \;\Delta D^{\dagger}D},
\end{eqnarray}
where $\Delta$ is assumed to be a positive constant for simplicity. It
can also contain derivatives. In such a case, the resulting free-part of
the dimeron Lagrangian contains derivatives.

We then make a change of variables,
\begin{eqnarray}
 D \!&=&\! 
  D'\left(
     1+a\phi^{\dagger}\phi+\cdots
    \right)
%  \nonumber \\
%  &&
  +\phi^{2}\!
  \left(
   b+c\phi^{\dagger}\phi+\cdots
  \right),
  \\
 D^\dagger \!&=&\!
  (D')^\dagger\left(
               1+a\phi^{\dagger}\phi+\cdots
              \right)
%  \nonumber \\
%  &&
  +(\phi^\dagger)^{2}\!
  \left(
   b+c\phi^{\dagger}\phi+\cdots
  \right),
  \nonumber \\
 &&
\end{eqnarray}
where $a$, $b$, and $c$ are real constants. Here and hereafter the
ellipses denote the terms of $\phi$'s and $\phi^\dagger$'s with
derivatives. Note that we assign $D$ and $D^\dagger$ the particle number
$-2$ and $+2$ respectively, with the assignment of the particle number
$-1$ to $\phi$.  We may include terms of higher powers in
$\phi^\dagger\phi$, $D$, and $D^\dagger$, but such terms do not
contribute to the three-body sector, and are thus ignored.

Under the change of variables, we have
\begin{eqnarray}
 \Delta D^{\dagger}D
  &=& 
  \Delta (D')^{\dagger}D'
  +b\Delta 
  \left(
   (D')^{\dagger} \phi^{2}+{\mbox h.c.}
  \right)
  \nonumber \\
 &&{}
  +2a\Delta  
  (D')^{\dagger}D' \phi^{\dagger}\phi
  \nonumber \\
 &&{}
  +\left(
    c+ab
   \right)
  \Delta
  \left(
   (D')^{\dagger}\phi^{\dagger}\phi^{3}
   +
   {\mbox h.c.}
  \right)
 \nonumber \\
 &&{}
  +b^{2}\Delta
  \left(
   \phi^{\dagger}\phi
  \right)^{2}
  +2bc\Delta
  \left(
   \phi^{\dagger}\phi
  \right)^{3}
  \nonumber \\
 &&{}+ \cdots.
\end{eqnarray}

This is not the whole story, however. One needs to consider the Jacobian
contribution unless one uses dimensional regularization,
\begin{equation}
 dD^\dagger dD = d{D'}^\dagger dD' J(\phi, \phi^\dagger)
\end{equation}
The Jacobian may be written as an action,
\begin{equation}
 J(\phi, \phi^\dagger)
 = \exp\left[-\int d^4x \; \mathcal{L}_{\rm Jacobian}\right],
\end{equation}
up to a numerical constant, with $\mathcal{L}_{\rm Jacobian}$ being expanded
in terms of local terms,
\begin{equation}
 \mathcal{L}_{\rm Jacobian} = 
  \delta M \phi^{\dagger}\phi 
  + Aa^2 \left(\phi^{\dagger}\phi\right)^2
  + Ba^3 \left(\phi^{\dagger}\phi\right)^3
  + \cdots,
  \label{JacobianL}
\end{equation}
with $\delta M$, $A$, and $B$ are regularization dependent real
constants. The calculation of the Jacobian must be consistent with the
regularization of the other parts, and it is difficult to ensure the
consistency. See Ref.~\cite{Harada:2009nb} for an example of consistent
calculation of a Jacobian factor in the lattice regularization. Here we
do not try to calculate it, since it does not affect the following
argument. 

The first term in Eq.~\eqref{JacobianL} may be absorbed in the
definition of the mass. With the other terms being included, the total
Lagrangian becomes
\begin{eqnarray}
 \mathcal{L}_{E}'&=&\mathcal{L}_{E} + 
  \Delta (D')^{\dagger}D'
  +b\Delta 
  \left(
   (D')^{\dagger} \phi^{2}+{\mbox h.c.}
  \right)
  \nonumber \\
 &&{}
  +2a\Delta  
  (D')^{\dagger}D' \phi^{\dagger}\phi
  \nonumber \\
 &&{}
  +\left(
    c+ab
   \right)
  \Delta
  \left(
   (D')^{\dagger}\phi^{\dagger}\phi^{3}
   +
   {\mbox h.c.}
  \right)
 \nonumber \\
 &&{}
  +\left(b^{2}\Delta+Aa^2\right)
  \left(
   \phi^{\dagger}\phi
  \right)^{2}
  +\left(2bc\Delta+Ba^3\right)
  \left(
   \phi^{\dagger}\phi
  \right)^{3}
  \nonumber \\
 &&{}+ \cdots.
\end{eqnarray}
Thus, to eliminate the terms $(\phi^\dagger\phi)^2$,
$(\phi^\dagger\phi)^3$, and $(D')^{\dagger}\phi^{\dagger}\phi^{3} +
{\mbox h.c.}$ in the whole Lagrangian, we choose $a$, $b$, and $c$ to
satisfy the following relations,
\begin{eqnarray}
 0&=&(c+ab)\Delta, \\
 0&=&-\frac{c_{0}}{4}+b^{2}\Delta+Aa^{2}, \\
 0&=&-\frac{d_{0}}{36}+2bc\Delta+Ba^{3}.
\end{eqnarray}
By eliminating $b$ and $c$ we have
\begin{eqnarray}
 \left(
  -\frac{1}{2}B
  -A
 \right)
 a^{3}
 +\frac{c_{0}}{4}a
 +\frac{d_{0}}{72}
 =0.
\end{eqnarray}
There is at least one real solution for $a$. Once the solution for $a$
is obtained, the solutions for $b$ and $c$ are obtained easily. We have
shown that, even in the presence of the contributions from the Jacobian,
the terms $(\phi^\dagger\phi)^2$, $(\phi^\dagger\phi)^3$, and
$(D')^{\dagger}\phi^{\dagger}\phi^{3} + {\mbox h.c.}$ can be eliminated
but, as we claimed before, the relations between the coupling constants
of the original theory and those in the theory with the dimeron are not
simple at all.

\section{An example of NOOL diagram contributions to the RGE for
 $d_{0}$}
\label{sec:appendix:NOOL}

\begin{figure}[h]
 \includegraphics{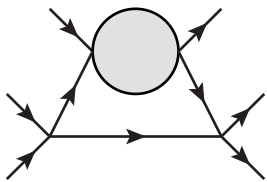}
 \caption{\label{type_a} An example of the NOOL diagram contributing to
 the RGE for $d_{0}$.}
\end{figure}

In this appendix, we illustrate how the NOOL diagram contributions are
evaluated by showing an example. 

%\comment{an example}

The amplitude of the diagram shown in Fig.~\ref{type_a} is given by
\begin{eqnarray}
 &&\int \frac{d^{4}k}{(2\pi)^{4}}
  \left(
   -ic_{0}-i4c_{2}\bfk^{2}-i\frac{b_{2}}{M}\bfk^{2}
  \right)^{2}
  \nonumber \\
 &&{}\times
  \left(-i{\cal A}
   \left(
    \mu, \frac{\bfk^{2}}{4},\frac{\bfk^{2}}{4}
   \right)
  \right)
  \nonumber \\
 & &\times
  \left[\frac{i}{k^{0}-\frac{\bfk^{2}}{2M}+i\epsilon}\right]^{2}
  \frac{i}{-k^{0}-\frac{\bfk^{2}}{2M}+i\epsilon}
  \nonumber \\
 &=&
  -
  \frac{i}{2\pi^{2}}
  \int^{\Lambda}_{0} dk 
  \left(
   c_{0}+4c_{2}k^{2}+\frac{b_{2}}{M}k^{2}
  \right)^{2}
  \nonumber \\
 &&{}\times
  {\cal A}
  \left(
   \frac{\sqrt{3}k}{2}, \frac{k^{2}}{4}, \frac{k^{2}}{4}
  \right)
  \frac{k^{2}}{\left[-k^{2}/M+i\epsilon\right]^{2}},
\end{eqnarray}
where the external energies and momenta are set to zero. The first
factor comes from the two two-body vertices. $\mathcal{A}$ stands for
the two-body amplitude~\eqref{offshell2bodyamp}, corresponding to the
blob in Fig.~\ref{type_a}. We have integrated over $k^0$ by picking up
the pole at $k^0=-k^2/2M +i\epsilon$ by the contour integration on the
upper half plane. Note that other poles and the cut in the amplitude is
on the lower half plane. The cutoff is introduced for the relative
momentum, in order to preserve the Galilean invariance of the
theory~\cite{Harada:2006cw}.

%\comment{shell modes}

The contribution to the RGE comes from the shell-mode part of the
integral, $\Lambda-d\Lambda <k< \Lambda$:

\begin{eqnarray}
 &&
  \frac{-i}{2\pi^{2}}\left(
    c_{0}+4c_{2}\Lambda^{2}+\frac{b_{2}}{M}\Lambda^{2}
   \right)^{2}
  {\cal A}
  \left(
   \frac{\sqrt{3}\Lambda}{2}, \frac{\Lambda^{2}}{4}, \frac{\Lambda^{2}}{4}
  \right)
  \nonumber \\
 &&{}\times
  \frac{\Lambda^{2}}{\left[-\Lambda^{2}/M+i\epsilon\right]^{2}}d\Lambda
  \nonumber \\
 &=&
 \frac{-iM^{2}}{2\pi^{2}\Lambda^{2}}\!
 \left(\!
  c_{0}+4c_{2}\Lambda^{2}+\frac{b_{2}}{M}\Lambda^{2}
  \!
 \right)^{2}\!\!
 {\cal A}
 \left(
  \frac{\sqrt{3}\Lambda}{2}, \frac{\Lambda^{2}}{4}, \frac{\Lambda^{2}}{4}
 \right)\!
 d\Lambda
 \nonumber \\
 &=&
  -4i
  \frac{2\pi^{2}}{\Lambda^{4}}
  \left(
   x+y+z
  \right)^{2}
  {\cal A}
 \left(
  \frac{\sqrt{3}\Lambda}{2}, \frac{\Lambda^{2}}{4}, \frac{\Lambda^{2}}{4}
 \right)d\Lambda.
\end{eqnarray}
We have the same contributions from the interchanges of the external
momenta, so that we multiply it with $3\times3$, obtaining
\begin{equation}
 -36i
  \frac{2\pi^{2}}{\Lambda^{4}}
  \left(
   x+y+z
  \right)^{2}
  {\cal A}
 \left(
  \frac{\sqrt{3}\Lambda}{2}, \frac{\Lambda^{2}}{4}, \frac{\Lambda^{2}}{4}
 \right)d\Lambda.
\end{equation}
The two-body amplitude in terms of $X$, $Y$, and $Z$ is given by
\begin{eqnarray}
 &&\mathcal{A}(\mu,\bfk_1^2,\bfk_2^2)
  \nonumber \\
 &=&
  \frac{4\pi^{2}}{M\Lambda}
  \bigg[
  1-\frac{\pi}{2}\frac{\mu}{\Lambda}
  +\frac{X^{2}}{Y}
  +\frac{\mu^{2}}{\Lambda^{2}}
  \bigg(
  \frac{X^{2}Z}{Y^{2}} +1
  \bigg)
  \nonumber \\
 &&{}
  -3
  \left(
   \frac{2\mu^{2}}{\Lambda^{2}}
   +
   \frac{\bfk_1^{2}+\bfk_{2}^{2}}{\Lambda^{2}}
  \right)
%   \nonumber \\
%  &&{}
%   \times
  \frac{X(X-1)}{Y}
  \left(
   \frac{X^2}{Y} + 1
  \right)
  \bigg]^{-1}.
  \nonumber \\
\end{eqnarray}
The combination $1+X^2/Y$ is written in terms of the two-body scattering
length, $a_2$, as
\begin{equation}
 1+ \frac{X^2}{Y} = \frac{\pi}{2}\frac{1}{a_{2}\Lambda},
  \label{oneovera2}
\end{equation}
and the combination $X^2Z/Y^2+1$ is written in terms of the two-body
effective range, $r_{e}$, as
\begin{equation}
 \frac{X^2Z}{Y^2}+1= \frac{\pi}{4}r_{e}\Lambda.
\end{equation}
Note that the cutoff $\Lambda$ is small enough so that the effective
range expansion is valid. If we denote the two-body cutoff
$\Lambda_{0}^{(2)}$, above which the effective theory description breaks
down, we assume
\begin{equation}
 \Lambda \ll \Lambda_{0}^{(2)}.
\end{equation}
We also assume that the effective range is of a natural size, i.e., 
\begin{equation}
 r_{e}\Lambda_{0}^{(2)} \sim \mathcal{O}(1),
\end{equation}
while the scattering length may (or may not) be fine-tuned. Thus, we
have
\begin{equation}
 r_{e}\Lambda \ll 1.
\end{equation}
The combination $3X(X-1)(X^2/Y+1)/Y$ is related to the off-shell parameter
$h$ through
\begin{equation}
 \frac{X(X-1)}{Y}\left(\frac{X^2}{Y}+1\right) 
  =\frac{\pi}{2}\frac{h\Lambda^2}{a_{2}\Lambda}.
  \label{hovera2}
\end{equation}
We assume that $h$ is also of a natural size,
\begin{equation}
 h\left(\Lambda_{0}^{(2)}\right)^2 \sim \mathcal{O}(1),
\end{equation}
thus
\begin{equation}
 h\Lambda^2 \ll 1.
\end{equation}
Therefore the quantity in Eq.~\eqref{hovera2} is much smaller than
the one in Eq.~\eqref{oneovera2}.

We may expand the corresponding terms in
$\mathcal{A}(\sqrt{3}\Lambda/2, \Lambda^2/4, \Lambda^2/4)$, keeping
$1+X^2/Y$ in the denominator. We finally get
\begin{equation}
 -72i
\frac{(2\pi^{2})^{2}}{M\Lambda^{5}}
\frac{V^{2}}{S^{2}}(T-6U)d\Lambda,
\end{equation}
where $S$, $T$, $U$, and $V$ are defined in Eqs.~\eqref{combinations}.

\section{Shell mode contribution of the two-loop diagram with three
 symmetric lines}
\label{sec:appendix:twoloop}

In this Appendix, we consider the shell mode contribution of the
two-loop diagram shown in the fifth diagram in Fig.~\ref{rge3}.

\begin{figure}
 \includegraphics[width=0.9\linewidth,clip]{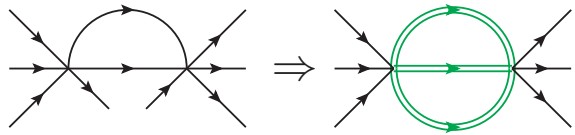}
 \caption{\label{tricky} The NOOL diagram, in which the identification
 of the shell mode contribution is not so obvious.}
\end{figure}

This NOOL diagram is obtained by contracting two legs of the one-loop
diagram as shown in Fig.~\ref{tricky}. The additional line connects the
same vertices as two lines in the original diagram do. As we explained
in Sec.~\ref{sec:normal-ordering}, the three lines should be treated in
a symmetric way and we should impose the cutoff so that Galilean
invariance is maintained. By doing so, the momenta of the three lines
are related to the shell mode.

A trouble is that there seems no obvious way to do so by just
``modifying'' the shell mode momentum assignment, as we did in
Fig.~\ref{shell-mode}. Fortunately, however, there is a trivial way to
identify the shell mode contribution of the diagram starting from the
amplitude,
\begin{eqnarray}
 I(\Lambda)&\equiv&
  \frac{1}{6}(-id_{0})^2
  \int\! \frac{d^{4}k}{(2\pi)^{4}}
  \int\! \frac{d^{4}l}{(2\pi)^{4}}
%   \int_{-\infty}^{+\infty} \frac{dk^0}{2\pi}
%   \int \frac{d^{3}\bfk}{(2\pi)^{3}}
%   \int_{-\infty}^{+\infty} \frac{dl^0}{2\pi}
%   \int \frac{d^{3}\bfl}{(2\pi)^{3}}
%   \nonumber \\
%  &&{}
%   \times
  \frac{i}{-k^{0}-\bfk^{2}/2M+i\epsilon}
  \nonumber \\
 &&{}
  \times
  \frac{i}{k^{0}/2+l^{0}-(\bfk/2+\bfl)^{2}/2M+i\epsilon}
  \nonumber \\
 &&{}
  \times
  \frac{i}{k^{0}/2-l^{0}-(\bfk/2-\bfl)^{2}/2M+i\epsilon},
\end{eqnarray}
where the external energy and momentum are set to zero for
simplicity. The domain of the momentum integrations is restricted to the
region $|\bfk| <\Lambda$ and $|\bfl|<\Lambda$. These cutoffs do not
break the Galilean invariance. Thus, the shell mode contribution is
identified as $I(\Lambda)-I(\Lambda-d\Lambda)$. A simple calculation
leads to
\begin{equation} 
 I(\Lambda)-I(\Lambda-d\Lambda)=
  6\frac{i(2\pi^{2})^{2}}{M\Lambda^{5}}cv^{2}d\Lambda.
\end{equation}

\section{Pole at $\xi=1/a$}
\label{sec:appendix:pole}

In this appendix, we demonstrate that the pole at $\xi=1/a$ appearing
in Eq.~\eqref{dvdt} is due to the existence of a two-body bound state.

In Appendix~\ref{sec:appendix:NOOL}, we have shown how the two-body
amplitude is embedded in the three-body NOOL diagrams and contributes to
the RGE for $v$. The two-body amplitude in the shell mode is given by 
\begin{equation}
 \mathcal{A}
  \left(
   \frac{\sqrt{3}\Lambda}{2},\frac{\Lambda^2}{4},\frac{\Lambda^2}{4}
  \right),
\end{equation}
and we have argued that the effective range and the off-shell parameter
terms can be expanded, so that effectively the amplitude appears as
\begin{equation}
 \mathcal{A}\sim \frac{8\pi}{M\Lambda}
  \left[
   \frac{1}{a_{2}\Lambda} -\frac{\mu}{\Lambda}
  \right]^{-1},
  \label{Asim}
\end{equation}
with $\mu=\sqrt{3}\Lambda/2$. In the usual off-shell amplitude, $\mu$ is
written as
\begin{equation}
 \mu=\sqrt{-MP^0+\bfP^2/4-i\epsilon},
\end{equation}
the value of $\mu$ corresponds to the energy
\begin{equation}
 P^0 = -\left(\frac{\sqrt{3}}{2}\right)^2\frac{\Lambda^2}{M},
  \label{p0}
\end{equation}
when the total momentum $\bfP$ is zero. On the other hand, because of
Eq.~\eqref{oneovera2},
\begin{equation}
 1+\frac{1}{\xi} =\frac{\pi}{2}\frac{1}{a_{2}\Lambda},
\end{equation}
$\xi=1/a$ corresponds to 
\begin{equation}
 \frac{1}{a_{2}\Lambda}= \frac{\sqrt{3}}{2}.
\end{equation}
Substituting it to Eq.~\eqref{p0}, we get
\begin{equation}
 P^0 =- \frac{1}{Ma_{2}^2},
\end{equation}
which is nothing but the energy of a two-body bound state in the present
approximation.

\begin{figure}
 \includegraphics[width=0.9\linewidth,clip]{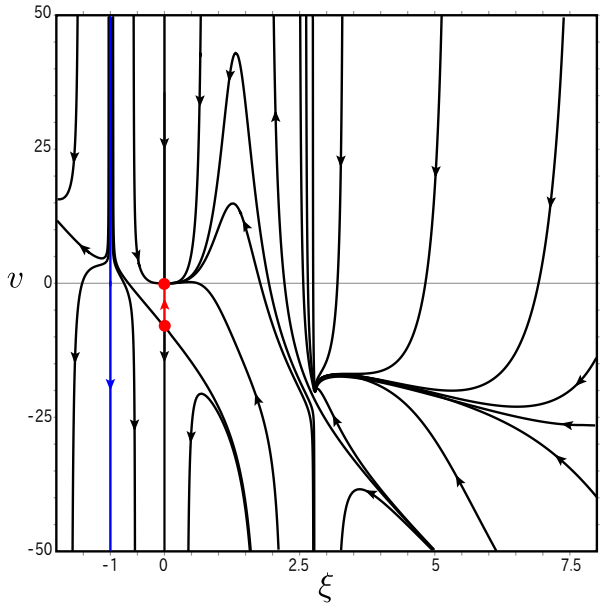}
 \caption{\label{hairpin} The RG flow in a wider region of the $\xi$-$v$
 plane. A ``wall'' is seen at $\xi=1/a$. Rapid change of $v$ near the
 ``wall'' is to cancel the enhancement due to the factor $1/(1-a\xi)$ in
 the RGE~\eqref{dvdt}.}
\end{figure}

In the two-body RGEs, no value of the cutoff hits the pole of the
amplitude, since $\mu$ is pure imaginary for any physical
energy-momentum. In the present case, the two-body amplitude is embedded
in the three-body diagrams, thus the ``total energy-momentum'' of the
two-body amplitude can take any real values so that $\mu$ can be real
and hits the pole of the amplitude Eq.~\eqref{Asim} as the value
$a_{2}\Lambda$ changes. When $\xi$ comes close to $1/a$, the three-body
shell-mode contributions become huge, and the value of $v$ changes
rapidly to cancel the cutoff dependence. Near the $\xi=1/a$, the
flows thus run almost vertically as shown in Fig.~\ref{hairpin}.

There is however a ``gate'' , $(1/a, -2/ab)$, at which the right hand
side of Eq.~\eqref{dvdt} is finite when $\xi$ approaches
$1/a$. Apparently no flow can pass the $\xi=1/a$ ``wall'' without
getting through the ``gate.''

\bibliography{../THREE,../NEFT,../EFT,../NPRG,../NLS}

\end{document}